\documentclass[usenatbib]{mn2e} 
\usepackage{graphicx,subfigure,times} 

\title[AGN heating in groups]{AGN heating in the centres of galaxy groups: a statistical study}
\author[N.N. Jetha et al]{N.N. Jetha$^{1,2}$\thanks{E-mail:
    nnj@star.sr.bham.ac.uk}, T.J. Ponman${^1}$, M.J. Hardcastle$^{3}$,
    and J.H. Croston$^{3}$\\$^{1}$School of Physics and Astronomy,
    University of Birmingham, Edgbaston, Birmingham B15
    2TT\\$^{2}$Service d'Astrophysique, CEA Saclay, L'Orme des
Merisiers, 91191 Gif-sur-Yvette, France\\$^{3}$School of Physics, Astronomy and Mathematics,
    University of Hertfordshire, College Lane, Hatfield, Hertfordshire
    AL10 9AB}

\begin{document}

\date{}

\maketitle

\


\newcommand{\Chandra}{\emph{Chandra}\ }
\newcommand{\Einstein}{\emph{Einstein}\ }
\newcommand{\ROSAT}{\emph{ROSAT}\ }
\newcommand{\XMM}{\emph{XMM-Newton}\ }
\newcommand{\MEKAL}{\textsc{MeKaL}\ }

\newcommand{\cm}{\mbox{\ensuremath{\mathrm{~cm}\ }}}
\newcommand{\m}{\mbox{\ensuremath{\mathrm{~m}\ }}}
\newcommand{\km}{\mbox{\ensuremath{\mathrm{~km}\ }}}
\newcommand{\pc}{\mbox{\ensuremath{\mathrm{~pc}\ }}}
\newcommand{\kpc}{\mbox{\ensuremath{\mathrm{~kpc}}}}
\newcommand{\Mpc}{\mbox{\ensuremath{\mathrm{~Mpc}}}}
\newcommand{\s}{\mbox{\ensuremath{\mathrm{~s}\ }}}
\newcommand{\ks}{\mbox{\ensuremath{\mathrm{~ks}\ }}}
\newcommand{\Gyr}{\mbox{\ensuremath{\mathrm{~Gyr}\ }}}

\newcommand{\keV}{\mbox{\ensuremath{\mathrm{~keV}\ }}}

\newcommand{\cmsq}{\ensuremath{\mathrm{\cm^2}\ }}
\newcommand{\cc}{\ensuremath{\mathrm{\cm^3}\ }}
\newcommand{\pcc}{\ensuremath{\mathrm{\cm^{-3}\ }}}
\newcommand{\pcmsq}{\mbox{\ensuremath{\mathrm{~cm^{-2}\ }}}}
\newcommand{\pMpc}{\ensuremath{\mathrm{\Mpc^{-1}\ }}}
\newcommand{\ps}{\ensuremath{\mathrm{\s^{-1}\ }}}
\newcommand{\pyr}{\ensuremath{\mathrm{\yr^{-1}\ }}}

\newcommand{\kmps}{\ensuremath{\mathrm{\km \ps}\ }}

\newcommand{\LCDM}{$\Lambda$CDM~}
\newcommand{\aj}{AJ}
\newcommand{\apjs}{ApJS}
\newcommand{\mnras}{MNRAS}
\newcommand{\apjl}{ApJL}
\newcommand{\aas}{A\&AS}
\newcommand{\apj}{ApJ}
\newcommand{\aap}{A\&A}
\newcommand{\aaps}{A\&AS}
\newcommand{\nat}{Nature}
\newcommand{\kmpspMpc}{\ensuremath{\mathrm{\km \ps \pMpc\,}}}
\newcommand{\ergps}{\ensuremath{\mathrm{\erg \ps}\ }}

\label{firstpage}

\begin{abstract}
We present gas temperature, density, entropy and cooling time profiles
for the cores of a sample of 15 galaxy groups observed with
{\it Chandra}.  We find that the entropy profiles follow a power-law
profile down to very small fractions of $R_{500}$.  Differences
between the gas profiles of groups with radio loud and radio quiet
BGGs are only marginally significant, and there is only a small
difference in the $L_X\!:T_X$ relations, for the central regions we
study with {\it Chandra}, between the radio-loud and radio-quiet objects in
our sample, in contrast to the much larger difference found on scales
of the whole group in earlier work.  However, there is evidence, from
splitting the sample based on the mass of the central black holes,
that repeated outbursts of AGN activity may have a long term
cumulative effect on the entropy profiles.  We argue that, to
first-order, energy injection from radio sources does not change the
global structure of the gas in the cores of groups, although it can
displace gas on a local level.  In most systems, it appears that AGN
energy injection serves primarily to counter the effects of radiative
cooling, rather than being responsible for the similarity breaking
between groups and clusters.

\end{abstract}

\begin{keywords}
galaxies: active - X-rays: galaxies: clusters
\end{keywords}

\section{Introduction}
\label{introduction}

Over the past decade cosmological simulations have become increasingly
sophisticated to the point where we can now follow the formation and
evolution of individual galaxy groups and clusters
\citep[e.g.][]{2001MNRAS.324..450T}.  Our knowledge about the
evolution of groups and clusters has also increased, due to a
combination of detailed X-ray studies using \ROSAT$\!$, \Chandra and \XMM and
deep, low frequency radio studies using the VLA radio
telescope (for example \citealt{2005ApJ...628..629N},
\citealt{2002ApJ...569L..79H}, \citealt{2000MNRAS.318L..65F}).

However, our understanding of galaxy groups and clusters is currently
impeded by two `heating problems'.  Whilst recent work has shown that
the dark matter distribution of galaxy clusters scales self-similarly
(e.g. \citealt{2005AdSpR..36..659P} and
\citealt{2006ApJ...640..691V}), the baryonic component (the hot gas)
does not follow self-similar scaling models (see for example,
\citealt{2003MNRAS.340..989S} and \citealt{2005AA...433..431P}).  This
appears to result from excess entropy in low mass clusters
(\citealt{1999Natur.397..135P} and \citealt{2003MNRAS.343..331P}),
which might be a consequence of feedback from supernova explosions or
active galactic nuclei (AGN). Secondly, the hot gas in cluster cores
is not cooling at the rates expected from the short cooling times
inferred \citep[e.g.][]{2001A&A...365L.104P}, suggesting that some
heating process must act to offset cooling. Since massive
early-type galaxies are found in the cores of almost all X-ray bright
groups and clusters, and are believed to host massive black holes, AGN
heating provides an attractive source for the required feedback
(e.g. \citealt{1995MNRAS.276..663B}, \citealt{2001MNRAS.325..676B},
\citealt{2001ASPC..240..363B}, \citealt{2001MNRAS.321L..20F}).

These models predict that a radio source could heat the IGM by doing
work on the hot gas both whilst the radio source is active, and once
the central AGN has switched off.  Models such as those described in
\citet{1998ApJ...501..126H} show that if a powerful radio source
erupts in a galaxy cluster, it will expand supersonically, heating the
IGM via shock heating.  A less powerful source in the early stages of
its evolution may show similar behaviour (see for example
\citealt{2001ApJ...549L.179R} for simulations and
\citealt{2003ApJ...592..129K} for observations of this phenomenon in
Centaurus A).  Further, once the radio bubbles are no longer being
`fed' by a central source, and begin to come into pressure equilibrium
with the IGM, additional work can be done (see for example
\citealt{2006astro.ph..2566N}).  Finally, the inflated bubble will
rise due to buoyancy, displacing gas further.  It is thought that if
repeated cycles of radio outbursts occur, and the effects are
integrated over time, then radio source heating may counteract
catastrophic cooling, even when no active source is present
\citep{2001ASPC..240..363B}.  

It has also been suggested that the energy input into the IGM
by a radio-loud AGN may also be responsible for similarity breaking in
groups and clusters (for example \citealt{1991ApJ...383..104K},
\citealt{1999A&A...347....1V}, and \citealt{2002MNRAS.333..145N}).
Occam's razor suggests that both similarity breaking and the
prevention of catastrophic cooling may be related, since both occur in
the same systems, and a similar source of energy injection is thought
to cause both.  The best place to look for evidence of such a
connection is probably not in rich clusters, but galaxy groups, since
the gas in the shallower potential wells of groups should be more
strongly affected by any embedded heat source, and it is in groups
that excess entropy is most clearly apparent.

The recent study of \citet{2005MNRAS.357..279C} found that the bulk X-ray
properties of galaxy groups are related to the presence of radio
galaxies within them; groups with radio-loud AGN tend to fall below
the $L_X\!:T_X$ relation, establishing a {\it prima facie} case for a
connection between AGN heating and group scaling properties. However,
since AGN radio outbursts are thought to be periodic, and the way in
which the intergalactic gas responds to the outbursts is poorly
understood, it is impossible to say whether there are long-lasting and
cumulative effects on the baryonic components of groups (such as would
be required to account for similarity breaking) or whether AGN
activity simply induces short-lived excursions in the $L_X:T_X$ plane
with no lasting effects on the temperature and entropy of the systems.
Further, the inference that it is the radio source that has an effect
on the IGM, rather than the conditions in the IGM being responsible
for triggering a radio source, may be incorrect; it could be that the
state of the IGM and the conditions required to trigger a radio source
form a feedback loop, such that the IGM attains a specific state,
which triggers a radio outburst, which in turn affects the IGM, and
eventually quenches the outburst, allowing the IGM to return back to
the initial state which triggered the outburst.

There appear to be three possible interpretations:
\begin{enumerate}
\item The radio source could modify the observed X-ray properties,
through heating or displacing gas.

\item Alternatively, the observed X-ray properties could be causing
the radio source by providing conditions that favour the triggering of a
radio outburst.  For example, the hot gas could provide a suitable
reservoir of gas for Bondi accretion
\citep[e.g.][]{2006astro.ph..2549A}, or the hot gas may have a steep
pressure gradient required to collimate large scale jets such as those
found in 3C31, in the group NGC~383 \citep{2002MNRAS.336.1161L}.

\item There is a third property of the system, such as BGG mass (see
Section~\ref{indsys}), or overall mass of the group, which correlates with
both the radio and X-ray properties, and is responsible for both.
\end{enumerate}

\begin{figure}
\scalebox{.4}{\includegraphics{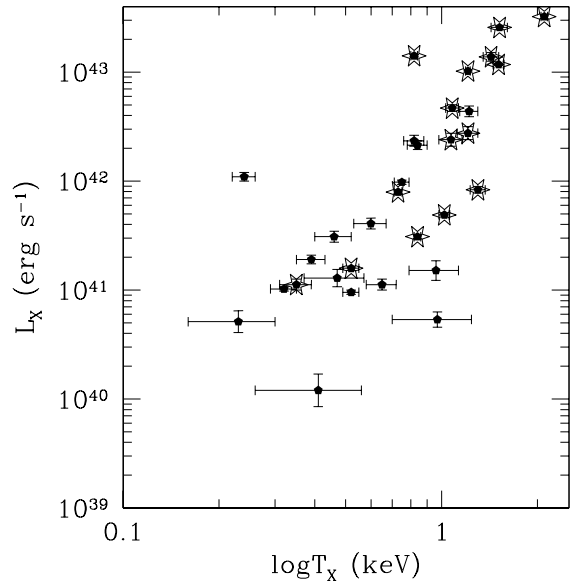}}
\caption{The \ROSAT $L_X:T_X$ relation of the groups used in the
Croston et al (2005) study.  The points marked with stars indicate the
groups analysed in this study; whilst higher luminosity systems are
favoured, this should not effect our conclusions (see text).}
\label{comparisonLX}
\end{figure}

In this paper, we use a sample of 15 galaxy groups to search for
evidence of radio source and hot gas interaction at the centres of
galaxy groups and to try to distinguish between the interpretations
discussed above.  The groups were taken from the GEMS sample of
\citet{OP2004}, and those groups in their G and H samples that had
good quality archival \Chandra data available were used for this study
(and correspond to a subset of the \citet{2005MNRAS.357..279C}
sample).  {\bf It should be noted that the parent GEMS sample is not,
in itself, unbiased, favouring systems with higher X-ray luminosities.
In this study, we have used a subset of the GEMS sample which had
available archival \Chandra data with observations longer than 30~ks;
thus we do not expect to explicitly bias our sample further.  However,
from Fig.~\ref{comparisonLX}, it can be seen that compared to the
sample of \citet{2005MNRAS.357..279C} our selection procedure does
tend to favour higher luminosity systems.  Since the X-ray luminosity
range in which the \ROSAT sample shows effects of heating is
nevertheless adequately sampled by our \Chandra sample, this slight
bias should not affect our conclusions.}

We present our sample and outline our data analysis in
Section~\ref{datanal}; we discuss our results in
Section~\ref{discussion}, and present our conclusions in
Section~\ref{conclusions}.  We assume that $H_0=72~\mathrm{km\ s^{-1}\
Mpc^{-1}}$ throughout.

\section{Data Analysis}
\label{datanal}
\subsection{Details of the sample and data reduction}

Details of the groups used in this study are given in
Table~\ref{chandraobs}.  Each \Chandra dataset was cleaned following
the procedure outlined in the {\sc ciao} threads on-line; the data
were reprocessed to apply the latest calibration products, and a new
events file was created.  Periods of high background were then removed
from the data by creating a light-curve of the background, and using a
3-$\sigma$ clipping algorithm to identify and remove times of high
background.  These datasets were then used in the subsequent spectral
and imaging analysis.

\begin{table*}
\caption{Details of the groups in this study, and the corresponding
\Chandra data.}
\begin{tabular}{lccccccl}
\hline source name & $z$& \multicolumn{2}{l}g$\!\!\!\!$roup centre co-ordinates&$R_{500}$&\Chandra obs& date observed & cleaned exposure\\
            &    &  $\alpha_{2000}$   &$\delta_{2000}$     &  (kpc) & number    &              &time         (ks)          \\ \hline        
NGC~383  &0.017 &01 07 25 &  32 24 46&510&2147&2000-11-06&42.8 \\
NGC~533  &0.019 &01 25 31 &  01 45 34&410&2880&2002-07-28&37.6  \\
NGC~720  &0.0058&01 53 01 & -13 44 20&260&492 &2000-10-12&39.6   \\
NGC~741 &0.019 &01 56 21 &  05 37 45&440&2223&2001-01-28&30.3  \\
NGC~1407 &0.0059&03 40 12 & -18 34 49&400&791 &2000-08-16&48.6  \\
NGC~3607 &0.0032&11 16 55 &  18 03 04&200&2073&2001-06-12&38.5   \\
NGC~4073 &0.020 &12 04 27 &  01 53 35&510&3234&2002-11-24&30.0   \\
NGC~4261 &0.0075&12 19 23 &  05 49 29&460&834 &2000-05-06&34.4   \\
NGC~4325 &0.026 &12 23 07 &  10 37 15&350&3232&2003-02-04&30.1    \\
NGC~4636 &0.0031&12 42 50 &  02 41 15&350&4415&2003-02-15&74.3     \\
HCG~62 &0.014 &12 53 06 & -09 12 14&490&921 &2000-01-25&47.4       \\
NGC~5044 &0.0090&13 15 24 & -16 23 09&440&3225&2002-06-07&83.1       \\
NGC~5171 &0.023 &13 29 22 &  11 44 06&410&3216&2002-12-10&34.7        \\
NGC~5846 &0.0057&15 06 29 &  01 36 21&320&788 &2000-05-24&29.9         \\
NGC~6338 &0.027 &17 15 23 &  57 24 41&620&4194&2003-09-17&47.3          \\
\hline
\end{tabular}
\vspace{0.2cm}
\begin{minipage}{16cm}
NOTES: $R_{500}$ is calculated using the \ROSAT temperatures of Osmond
and Ponman (2004), and the scaling relation of $R_{500}=391\times
T^{0.63}$ \citep{2005MNRAS.363..675W}. 
\end{minipage}
\label{chandraobs}
\end{table*}

\begin{table*}
\caption{Radio powers, galactic velocity dispersions, $\sigma$, and K-band
magnitudes, $M_K$ for the sources in our sample}
\begin{tabular}{llllcll}
\hline Source & 1.4~GHz radio power & radio flux &size of
radio source&$\sigma$ &$\sigma$ & $M_K$ \\
&$\mathrm{W\ Hz^{-1}}$               &reference   &at 1.4~GHz(kpc)&\kmps  &reference&  mag.\\ \hline    
NGC~383       &$3.2\times 10^{24}$  &1& 360&288.0&3&-23.57\\
NGC~533       &$2.2\times 10^{22}$  &1& 20 &224.0&4&-25.84\\
NGC~720       &$<1.0\times 10^{20}$ &2& -  &273.0&3&-24.57\\
NGC~741       &$7.9\times 10^{23}$  &1& 150&270.0&5&-26.13\\
NGC~1407      &$6.6\times 10^{21}$  &1& 10 &259.7&6&-25.15\\
NGC~3607      &$ 1.2\times 10^{20}$ &1& 4  &240.0&7&-23.49\\
NGC~4073      &$<3.9\times 10^{20}$ &2& -  &276.0&8&-26.04                \\
NGC~4261      &$2.3\times 10^{24}$  &1& 80 &316.0&9&-25.10                 \\
NGC~4325      &$<6.3\times 10^{20}$ &2& -  &-    &&-24.77                   \\
NGC~4636      &$6.1\times 10^{21}$  &1& 3  &208.0&9& -23.97       \\
HCG~62        &$3.2\times 10^{20}$  &2& 15 &251.0&10&  -25.09            \\
NGC~5044      &$6.2\times 10^{21}$  &1& 15 &240.0&11& -25.08               \\
NGC~5171      &$<2.6\times 10^{21}$ &2& -  &-    & & -24.73                \\
NGC~5846      &$1.5\times 10^{21}$  &2& 1  &261.0&9&-24.93                  \\
NGC~6338      &$9.5\times 10^{22}$  &2& 6  &347.0&12& -25.84                 \\
\hline
\end{tabular}
\vspace{0.2cm}
\begin{minipage}{16cm}
NOTES:
Column 3 gives the reference for the radio flux density:\\
1 Radio flux from NED.\\
2 Radio flux from NVSS \citep{1998AJ....115.1693C}.\\
Column 6 gives the reference for galactic velocity dispersion:\\ 
3 \citet{2000MNRAS.313..469S}\\
4 \citet{1999A&AS..140..327M}\\
5 \citet{1995A&A...297...28B}\\
6 \citet{2002A&A...395..431B}\\
7 \citet{2002MNRAS.333..517P}\\
8 \citet{1995ApJ...438..539F}\\
9 \citet{2002AJ....123.2990B}\\
10 \citet{1998yCat..33300423R}\\
11 \citet{1993MNRAS.265..553C}\\
12 \citet{1999MNRAS.305..259W}\\
K-band magnitudes collated from 2MASS
\end{minipage}
\label{physprops}
\end{table*}

\subsection{Imaging}
\label{imaging}
For each group, the cleaned data were smoothed using the adaptive
smoothing algorithm {\sc csmooth} which scales the size of the
smoothing Gaussian function until a pre-defined number of counts,
which corresponds to a user specified significance, is met.  This
implies that only structure which is `real' in a statistical sense is
smoothed. 

\begin{figure*}
\subfigure{\scalebox{.25}{\includegraphics{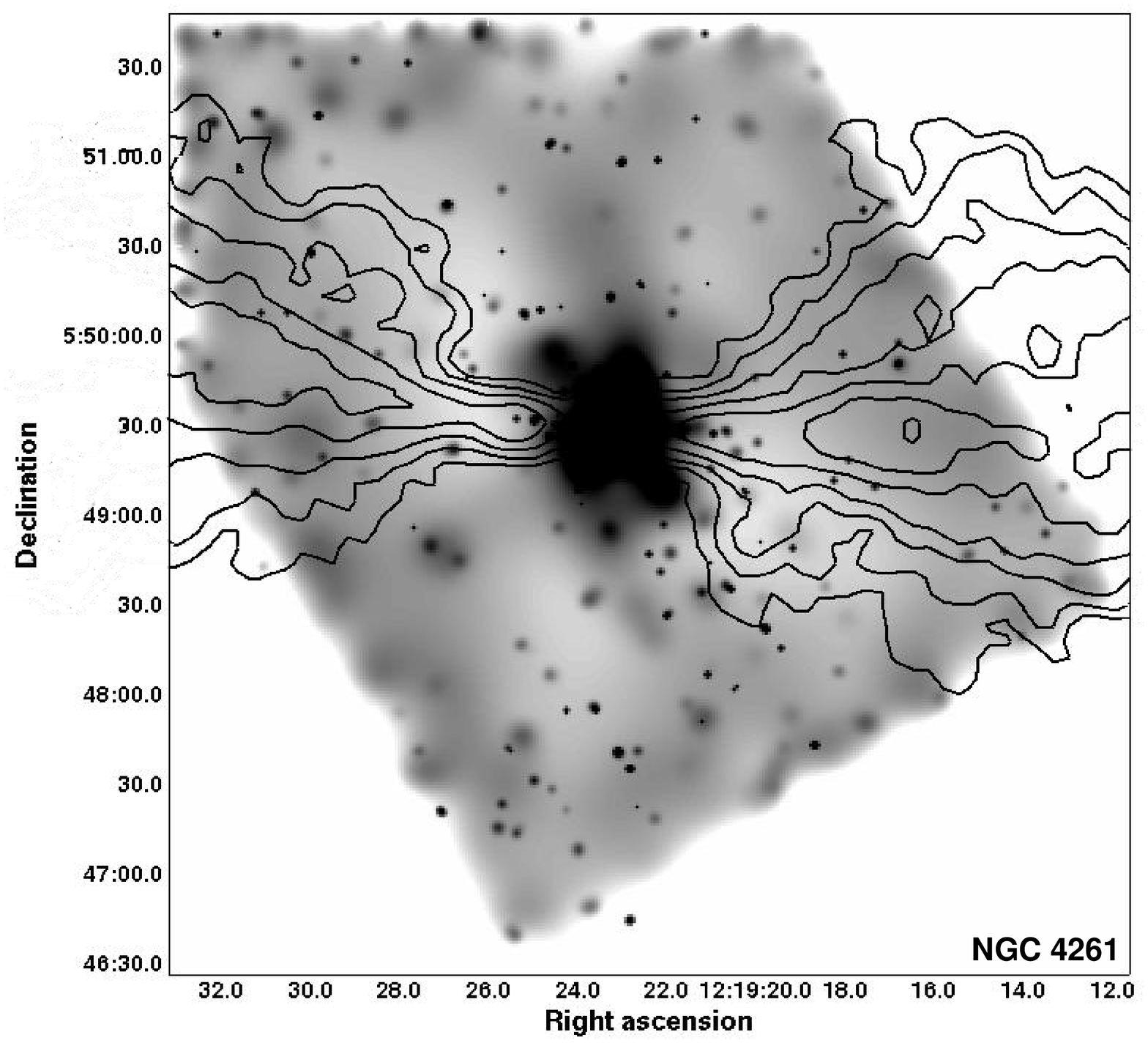}}}\\
\subfigure{\scalebox{.3}{\includegraphics{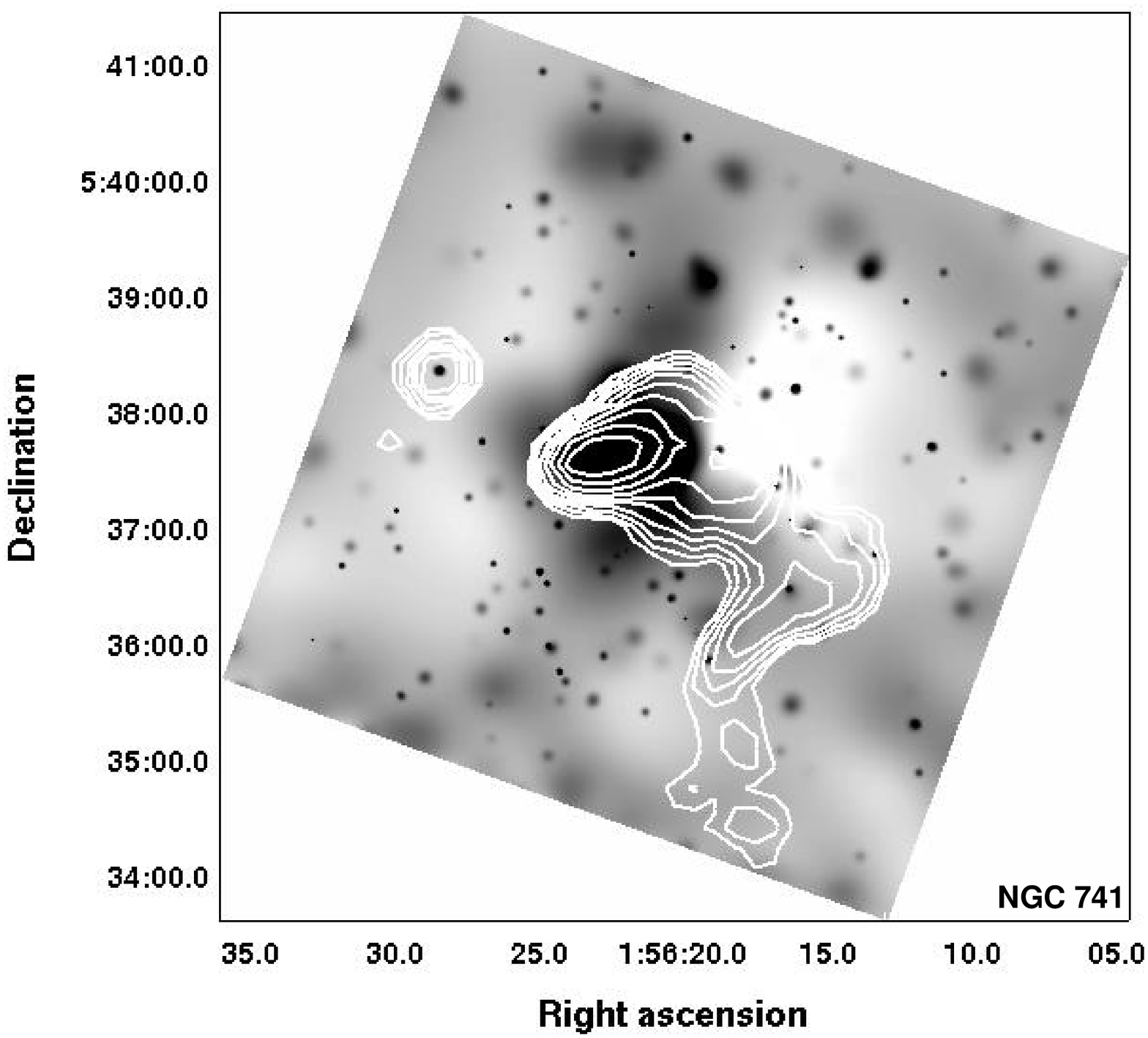}}}
\subfigure{\scalebox{.35}{\includegraphics{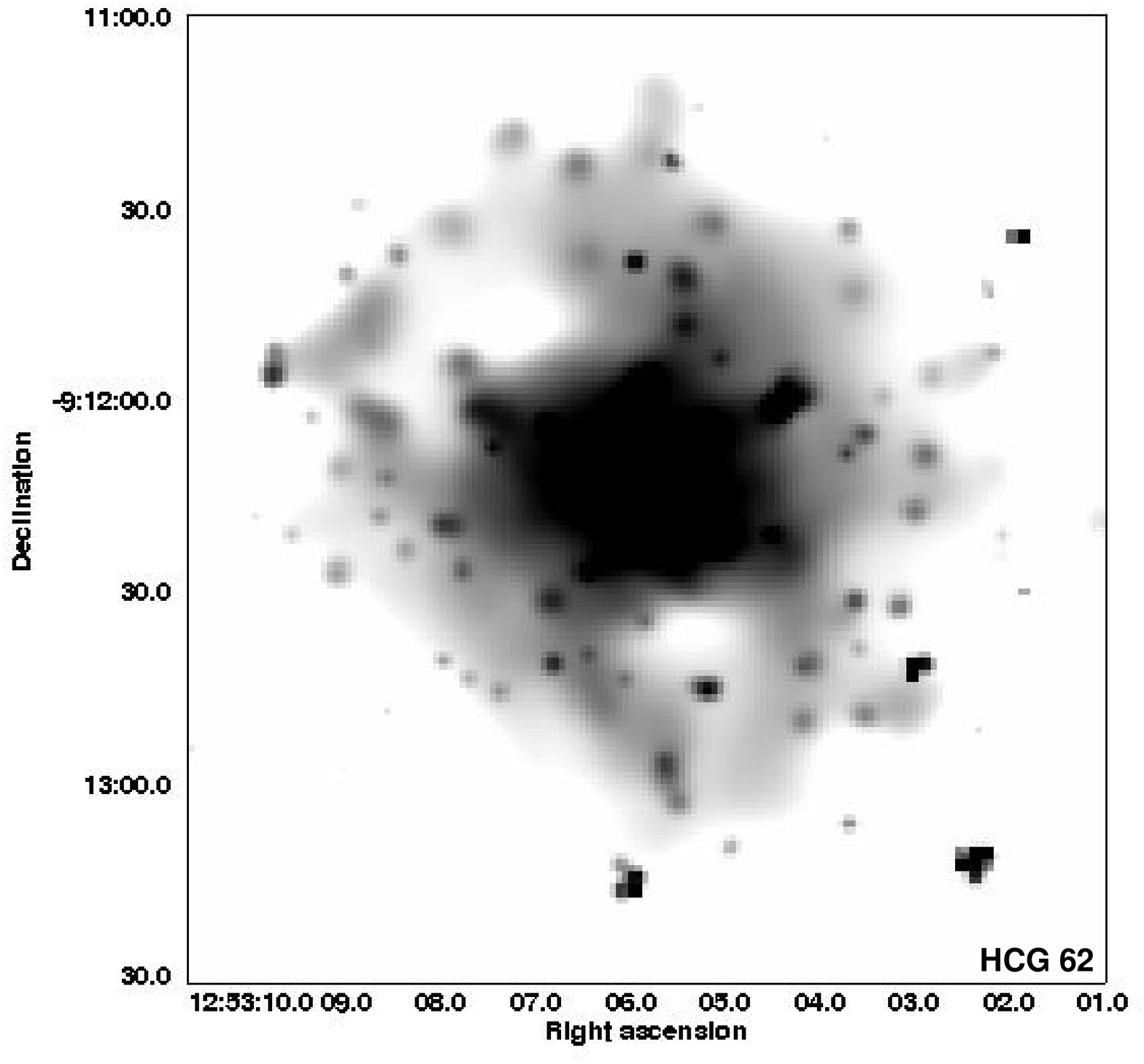}}}
\caption{Groups in our sample which exhibit spatial evidence of
interaction between the hot IGM and the AGN.  NGC~4261
shows filaments of hot gas outlining the radio lobes, whereas NGC~741
and HCG~62 exhibit ghost cavities, where there is no clear
correspondence between AGN emission (if any) and features in the
X-ray.}
\label{chandraimages}
\end{figure*}

From the images, there does appear to be evidence of disturbed hot gas
in some group cores; groups such as NGC~741 (Fig.~\ref{chandraimages})
and HCG~62 \citep{2004ApJ...607..800B} show `ghost' X-ray cavities,
and NGC~4261 shows `filaments' of hot gas outlining the radio lobes
(shown in Fig.~\ref{chandraimages}), which also extend to larger
scales \citep[see][]{2005MNRAS.357..279C}.  If the hot gas shows
evidence of spatial interaction with the radio plasma injected into
the group by the AGN, then it may be the case that a change in radial
temperature, density, or entropy distributions could be seen if the
AGN is injecting extra energy into the group.

\subsection{Spectral analysis}
\label{spectralan}
\subsubsection{Obtaining temperatures and luminosities}
\label{obtaininglt}

\begin{figure}
\subfigure{\scalebox{.3}{\includegraphics{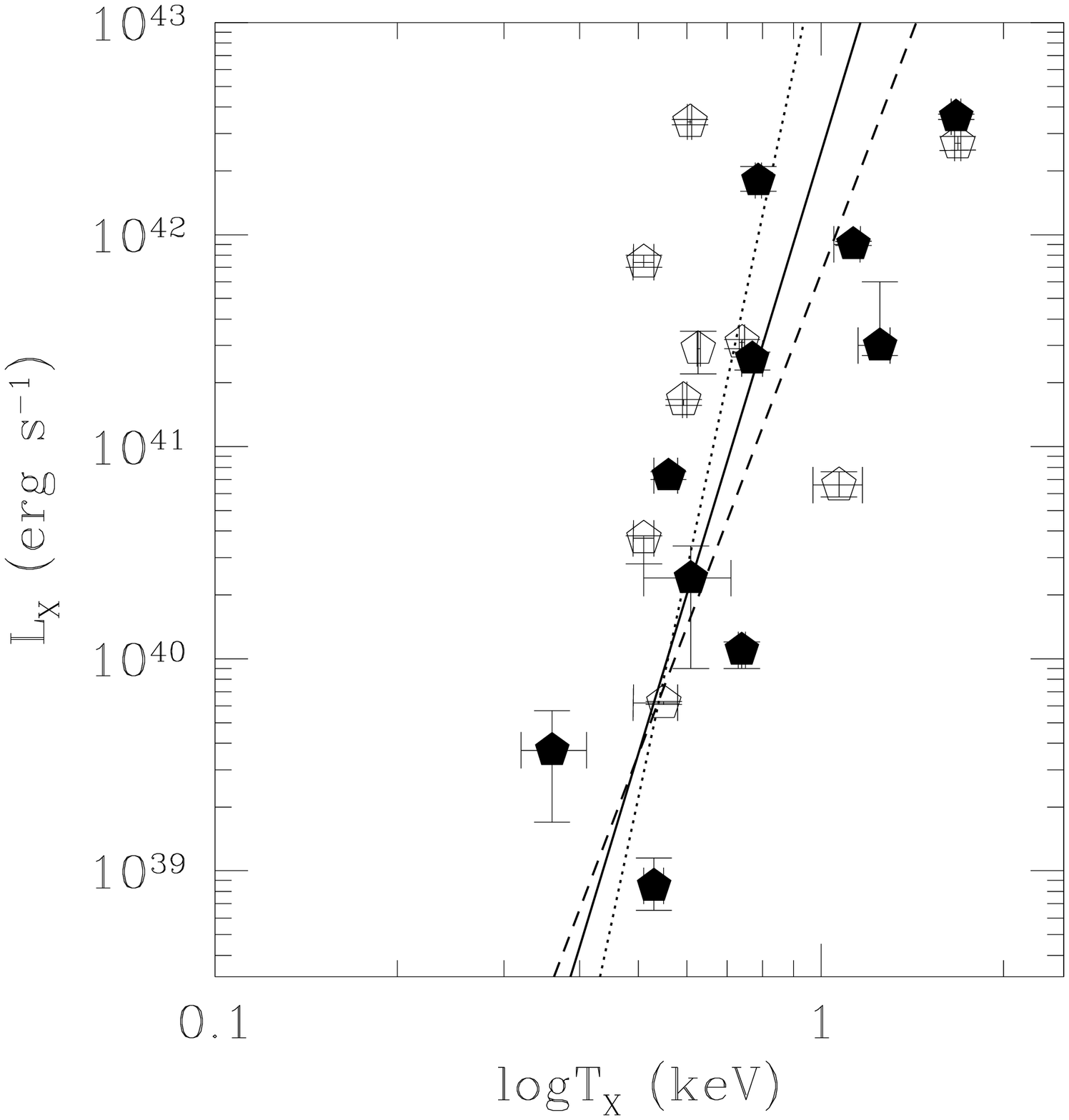}}}
\subfigure{\scalebox{.3}{\includegraphics{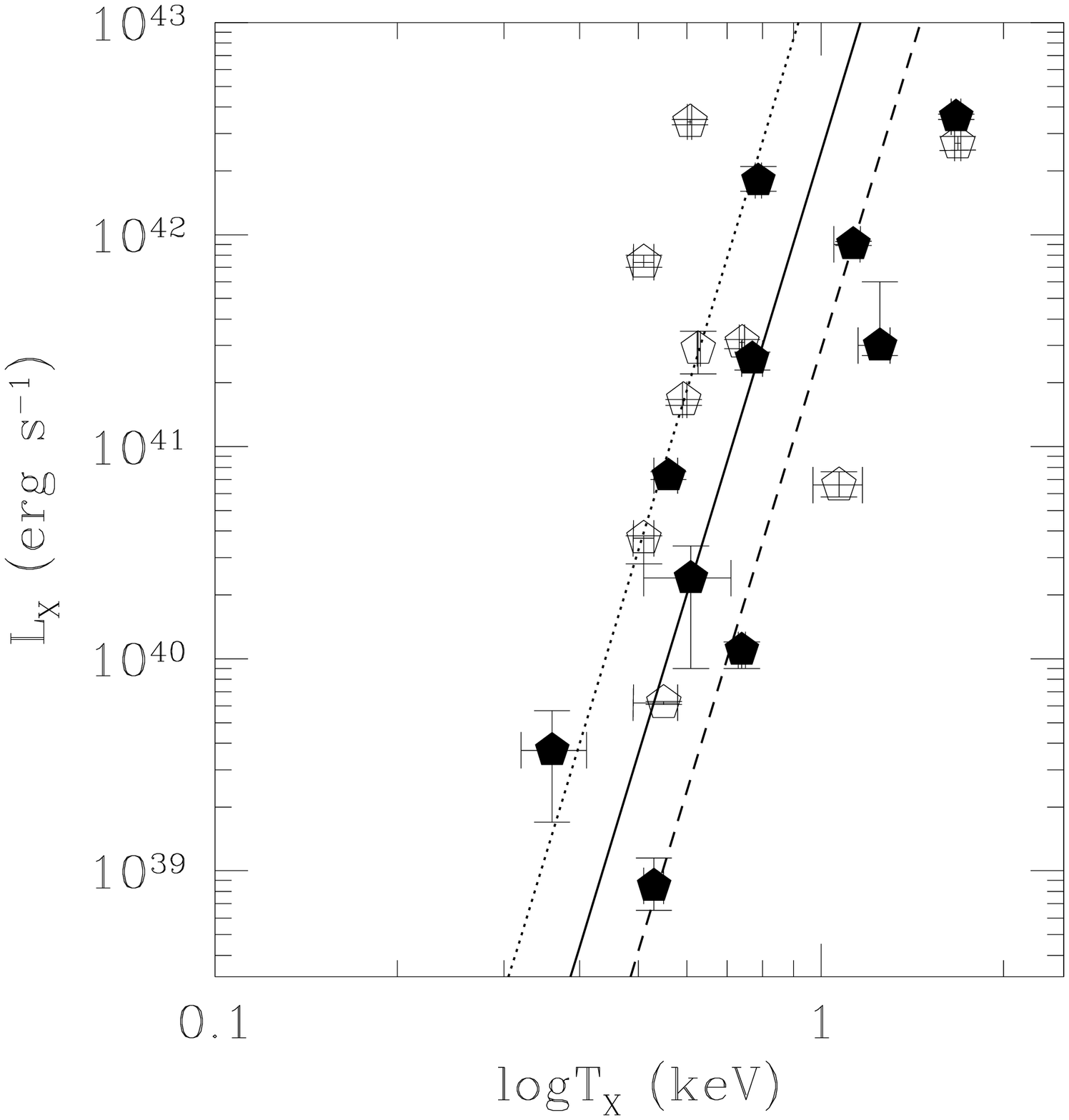}}}
\caption{The $L_X:T_X$ relation for the groups in our sample, using
\Chandra data.  Temperatures and luminosities were extracted in
regions that extended out to $0.05R_{500}$.  The solid points indicate
the radio loud groups and the open points the radio quiet groups.  We
define radio loud groups as groups whose BGG has
$\log(L_{1400})\geq21.5$.  In both panels, the solid line is the fit to the $L_X:T_X$
relation for the entire sample (see Section~\ref{LxTx}), the dotted
line is the fit to the radio quiet sample and the dashed line is the
fit to the radio loud sample. In the top panel, we fit gradients and
intercepts for all three plotted relations, whereas in the bottom
panel, we assume that the gradient of the relation for the entire
sample holds for both the radio quiet and the radio loud sample, and
fit only the intercepts.}
\label{LTrel}
\end{figure}

In order to investigate the effect of radio outbursts on the
$L_X$:$T_X$ relation for the inner regions of galaxy groups, spectra
were extracted for the systems in the sample out to a fiducial radius
of $0.05R_{500}$.  {\bf This radius was used as we had data for all
groups in our sample, apart from NGC~4261 which was excluded from the
$L_X$:$T_X$ analysis since the original X-ray observation did not have
data out to $0.05R_{500}$.  Our measurement of $R_{500}$ was obtained
using the $R_{500}:T_X$ relation of \citet{2005MNRAS.363..675W}, who
use the {\it XMM-LSS} for a sample of galaxy groups and clusters.}  In
addition to the groups shown in Table~\ref{chandraobs}, we also
obtained temperatures and luminosities for a further 4 groups
(NGC~1587, NGC~3665, NGC~3923, and IC~1459) that were not included in
the original sample due to the quality of the data (and hence did not
have a deprojection analysis done, see Section~\ref{deproj}).  Spectra
were extracted using the {\it CIAO} {\it acisspec} script, excluding point
sources but not the central AGN where one was present, and using a
local background immediately outside the region from where the
spectrum was extracted.  The spectra were fitted in {\sc xspec} with
an absorbed \MEKAL model, and a power law for the AGN component (if
present) which was left free to fit the data, and a power law with a
fixed index of 1.72 to account for any contribution from X-ray
binaries.  A temperature and unabsorbed luminosity within
$0.05R_{500}$ was measured for each group.  The results are shown in
Fig~\ref{LTrel}.

\subsubsection{Deprojection}
\label{deproj}
If the radio source in a specific group is having an effect on the
bulk properties of the hot gas, rather than simply displacing gas and
causing local perturbations, then it might be expected that an effect
would be seen in the overall radial properties of the group.  If radio
source heating is occurring, then elevated temperatures and excess
entropy may be present in those groups with current activity, compared to
those which are currently radio quiet.  By examining the radial
profiles of the gas properties, and comparing the profiles of
radio-loud groups with those of radio quiet groups, differences in
temperature and entropy may be observed.

For these reasons, we analysed the galaxy groups using an onion skin
deprojection method to obtain radially deprojected temperature
profiles.  In doing the onion-skin deprojection, we assumed that the
group was spherically symmetric, and that the gas had a filling factor
of unity.  Spectra were extracted in concentric annuli centred on the
brightest group galaxy (BGG), with the outermost annulus being used as
a local background.  The width of each annulus was determined by the
number of counts in each annulus; to do our spectral analysis, a
minimum of 1000 counts was required in each annulus.  We then followed
the deprojection method described in \citet{Jethaa}, fitting each spectrum with
a \MEKAL model to obtain temperatures and normalizations for each
shell.

The normalization of the \MEKAL model ($N_{mek}$) is related to the
density of the gas in the spherical shell by
\begin{equation}
N_{mek} = \frac{10^{-14}}{4\pi \left[D_A\left(1+z\right)\right]^2}\int n_e n_p dV,
\end{equation}
where $n_e$ and $n_p$ are the electron and proton densities of the
gas and $D_A$ is the angular size distance to the source.  Further
assuming that $n_p=1.18n_e$, and that the gas to be fitted occupies a
volume $V$, then the density of the gas is
\begin{equation}
n_e = \left\{\frac{4\pi\left[D_A\left(1+z\right)\right]^2 N_{mek}}{1.18\times10^{-14}V}\right\}^{\frac{1}{2}}.
\end{equation}

The density and temperature profiles obtained from the deprojection
were then used to calculate entropy, pressure and cooling time
profiles for the sources.  We calculate the entropy index, $S(r)$ for
a radial bin using:
\begin{equation}
S(r)=T(r)n(r)^{-2/3},
\label{entropy}
\end{equation} where $T$ is the temperature (in K) obtained for the
bin, and $n$ is the corresponding total density (electron density plus
proton density) for the bin.  Here $S$ is not a measure of the true
entropy, but is rather the adiabatic constant -- the ratio of the heat
capacity for heating at constant pressure to the heat capacity for
heating at constant volume ($C_p/C_V$), and is linked to the entropy
of the gas ($S^{\prime}$) by
\begin{equation}S=\frac{h^2}{2\pi\left(\mu\m_H\right)^{8/3}}\exp\left(S^{\prime}/c_v
- 5/3\right),\end{equation} where $h$ is the Hubble constant scaled to
$H_0\!=100$\kmpspMpc$\!$, $\mu$ is the reduced mass of hydrogen, $m_H$ is
the mass of a hydrogen atom, and $c_v$ is the specific heat capacity
for expansion at constant volume \citep[see][and references
therein]{1999MNRAS.307..463B}.  Thus, $S$ provides the most direct
observational indicator of gas entropy.

Pressure profiles $P(r)$ are generated similarly using
\begin{equation}
P(r)=\left[2.98\times10^{-10}\mathrm{Pa\ keV^{-1}\
cm^{3}}\right]n_{\mathrm{p}}(r)T(r),
\end{equation} In calculating the cooling time, $T_{cool}$, we note
that as most of the groups have temperatures $<$2\keV, the thermal
bremsstrahlung approximation is not valid.  Instead, an estimate of
the cooling time can be made by dividing the total internal energy of
the gas in a given radial bin
\begin{equation}
U(r)=\frac{3}{2}(n_e+n_p)VT(r),\end{equation} by the bolometric luminosity,
$L_{bol}$, obtained from {\sc xspec} when fitting the data.

\begin{figure}
\scalebox{.3}{\includegraphics{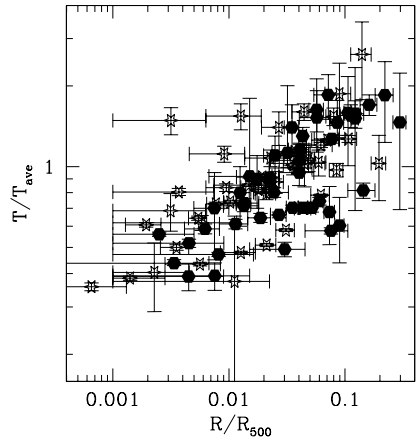}}
\caption{X-ray temperature profiles for the groups in the GEMS sample which
  have \Chandra data. Groups containing a BGG with $\log L_{1400}\geq
21.5$ are plotted with closed points, and groups with BGGs with a
radio power less then this are plotted with open points.}
\label{scaledtemp}
\end{figure}

\begin{figure}
\scalebox{.3}{\includegraphics{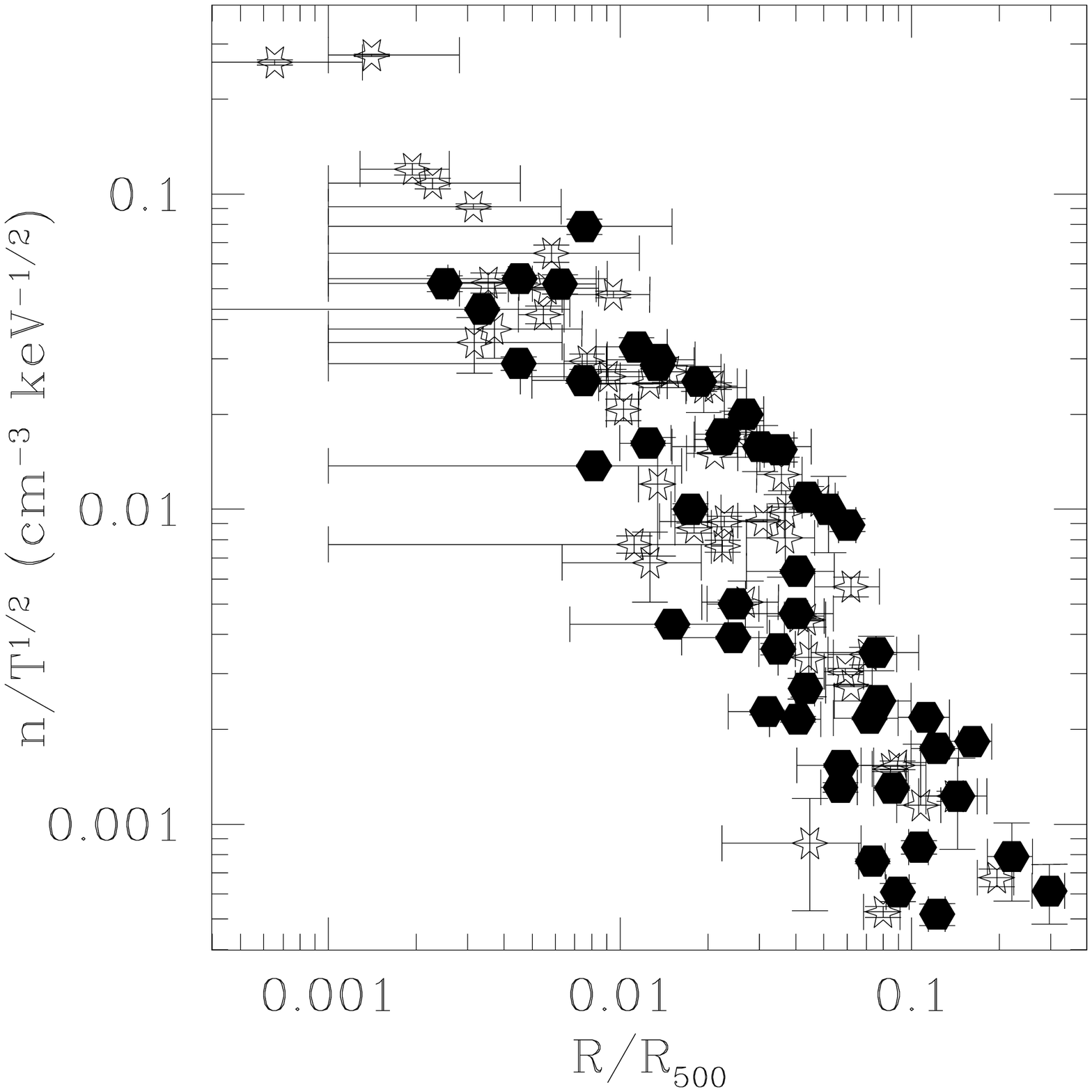}}
\caption{Gas density profiles for the same groups as in
  Fig~\ref{scaledtemp}.  The data are scaled horizontally as in
  Fig~\ref{scaledtemp} and vertically by $T^{1/2}$.}
\label{scaleddensity}
\end{figure}

\begin{figure}
\scalebox{.3}{\includegraphics{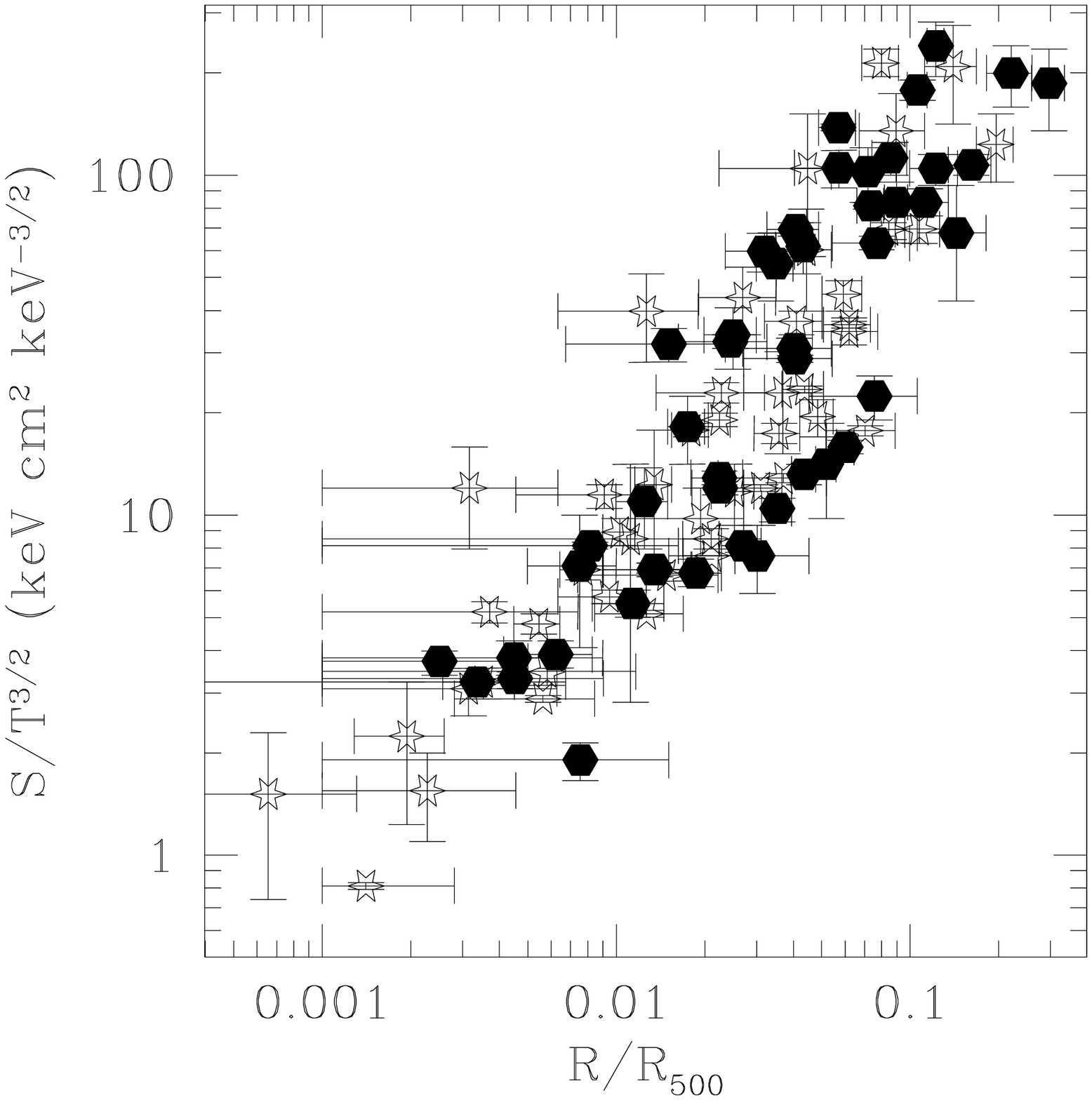}}
\caption{Gas entropy profiles for the same groups as in
  Fig~\ref{scaledtemp}.  The data are scaled horizontally as in
  Fig~\ref{scaledtemp} and vertically by $T^{2/3}$.}
\label{scaledentropy}
\end{figure}

\begin{figure}
\scalebox{.3}{\includegraphics{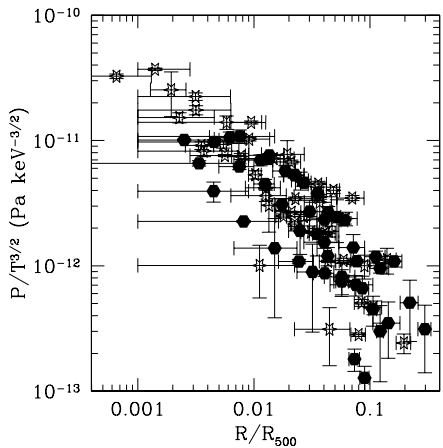}}
\caption{Gas pressure profiles for the same groups as in
  Fig~\ref{scaledtemp}.  The data are scaled horizontally as in
  Fig~\ref{scaledtemp} and vertically by $T^{3/2}$.}
\label{scaledpressure}
\end{figure}

\begin{figure}
\scalebox{.3}{\includegraphics{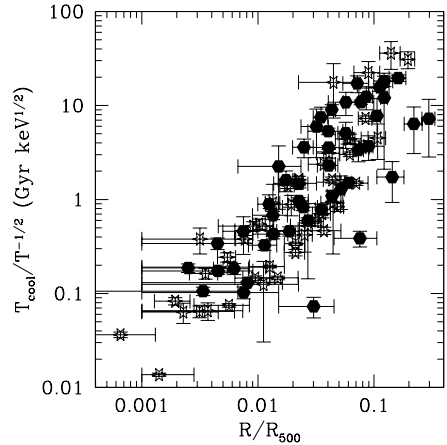}}
\caption{Gas cooling time profiles for the same groups as in
  Fig~\ref{scaledtemp}.  The data are scaled horizontally as in
  Fig~\ref{scaledtemp} and vertically unscaled.}  
\label{ct}
\end{figure}

However, since the groups are all of different masses, a meaningful
comparison of the derived quantities cannot be made without scaling to
take the mass differences into account.  Here, we use modified
self-similar scaling \citep{2003ApJ...594L..75V} such that radial
distances are scaled by $R_{500}$ (the radius at which the density of
the system is 500 times the critical density of the universe),
temperatures are scaled by $T_{ave}$, the average temperature of the
system (as obtained from the \ROSAT data of \citep{OP2004} with cool
cores excluded from the spectral analysis), density by
$T_{ave}^{1/2}$, entropy by $T_{ave}^{2/3}$, pressure by
$T_{ave}^{3/2}$.  We do not scale the cooling time profiles since in
the regime where $T<$2~keV X-ray emission is dominated by line
emission which is a function of temperature, which makes scaling by a
single temperature unrealistic.  The scaled
profiles are shown in Figs~\ref{scaledtemp}-\ref{ct}.

The scaled temperature, density and entropy profiles are then fitted
with power laws using an orthogonal regression algorithm as
implemented in the {\sc slopes} package \citep{1996ApJ...470..706A}.
The results of the fitting are shown in Table~\ref{fits}.

\begin{table}
\caption{Results of power law fits to our scaled temperature, density
and entropy profiles.  {\bf The normalisations represent the
logarithmic values for temperature, density and entropy of a 1~keV
group at $R_{500}$, assuming that the power law remains unbroken to
this distance.}} 
\begin{tabular}{lll}
\hline Quantity& Index        & Normalisation\\ \hline
Temperature (keV)    & 0.26$\pm$0.05&0.36$\pm$0.05  \\
Density ($\mathrm{cm^{-3}}$)        &-1.23$\pm$0.06&-4.1$\pm$0.1    \\
Entropy ($\mathrm{keV\ cm^2}$)& 1.08$\pm$0.05&3.05$\pm$0.10 \\ \hline
\end{tabular} 
 \vspace{0.2cm}
\begin{minipage}{8cm}
\small NOTES: The sub-samples were fitted in logarithmic space using
an orthogonal regression method.  Errors are 1$\sigma$ for two
interesting parameters.
\end{minipage}
\label{fits}
\end{table}

\section{Results}
\label{discussion}
\subsection{General observations}
Examining the scaled profiles, it appears that groups follow modified
self-similar scaling, as discussed by
\citet{2003ApJ...594L..75V}.  Looking closely at the entropy profiles,
it is apparent that there is no isentropic core, even down to small
fractions of the virial radius; the entropy profiles follow
predictions from models of cooling of gas without any significant
heating to raise the central entropy.  This is in contrast
to work probing hotter galaxy clusters, e.g. that of
\citet{2005ApJ...634..955V} who find isentropic cores within 10~\kpc
($\sim 0.01R_{500}$) in 2.2-5.5~\keV clusters.  The profiles presented
in this paper probe the hot gas in groups to similar fractions of
$R_{500}$, but no evidence of isentropic cores is seen.
Sanderson et al (in prep) also find a similar result to that presented
here for galaxy clusters observed with \Chandra$\!$.

\subsection{Radio loud versus radio quiet sources} 
\label{AGNsplit}
The observed profiles are of a mixture of groups with and without
current radio activity in the BGG.  If radio outbursts are injecting
energy into the IGM, then we might expect these outbursts to have an
effect on the gas properties of the IGM.  In order to investigate
this, we divide our sample into two, based on the 1.4~GHz radio
luminosity ($\log\left[L_{1400}/\left({\mathrm{W\
Hz^{-1}}}\right)\right]$) of the central AGN (given in
Table~\ref{physprops}).  We split our sample at three different radio
luminosities, $\log\left(L_{1400}\right)=$21, 21.5, and 22, in order
to establish if there are any differences as radio power increases.
The scaled density, temperature, and entropy profiles for each sample
are then fitted with power laws as described in
Section~\ref{spectralan}.

\begin{table*}
\caption{Fits to our sample split according to the radio power of the BGG.}
\begin{tabular}{lllllll}
\hline Sub-sample        & \multicolumn{2}{c}{Temperature}  & \multicolumn{2}{c}{Density}     & \multicolumn{2}{c}{Entropy}\\
              & gradient & intercept             & gradient &
intercept     &  gradient & intercept\\ \hline
$L_{1400}<21$       &0.22$\pm$0.07&0.3$\pm$0.1&-1.2$\pm$0.1&-4.1$\pm$0.1&1.1$\pm$0.1&3.0$\pm$0.2\\
$L_{1400}\geq 21$   &0.28$\pm$0.03&0.38$\pm$0.05&-1.23$\pm$0.07&-4.1$\pm$0.1&1.09$\pm$0.07&3.1$\pm$0.1 \\
                &&&&&&\\
$L_{1400}<21.5$     &0.24$\pm$0.04&0.38$\pm$0.06&-1.16$\pm$0.07&-3.9$\pm$0.1&1.00$\pm$0.08&2.9$\pm$0.2 \\
$L_{1400}\geq 21.5$ &0.30$\pm$0.01&0.42$\pm$0.02&-1.08$\pm$0.09&-3.7$\pm$0.2&1.04$\pm$0.05&3.02$\pm$0.09 \\
		&&&&&&\\
$L_{1400}<22$       &0.23$\pm$0.04&0.33$\pm$0.07&-1.17$\pm$0.07&-3.9$\pm$0.1&1.01$\pm$0.08&2.9$\pm$0.1 \\
$L_{1400}\geq22$
&0.34$\pm$0.04&0.45$\pm$0.06&-1.3$\pm$0.1&-4.2$\pm$0.2&1.21$\pm$0.09&3.2$\pm$0.1
\\ \hline
\end{tabular}
\vspace{0.2cm}
\begin{minipage}{16cm}
\small NOTES: The sub-samples were fitted in logarithmic space using
an orthogonal regression method.  Errors are 1$\sigma$ for two
interesting parameters.
\end{minipage}
\label{results_fitsAGN}
\end{table*}

\begin{figure}
\subfigure{\scalebox{.3}{\includegraphics{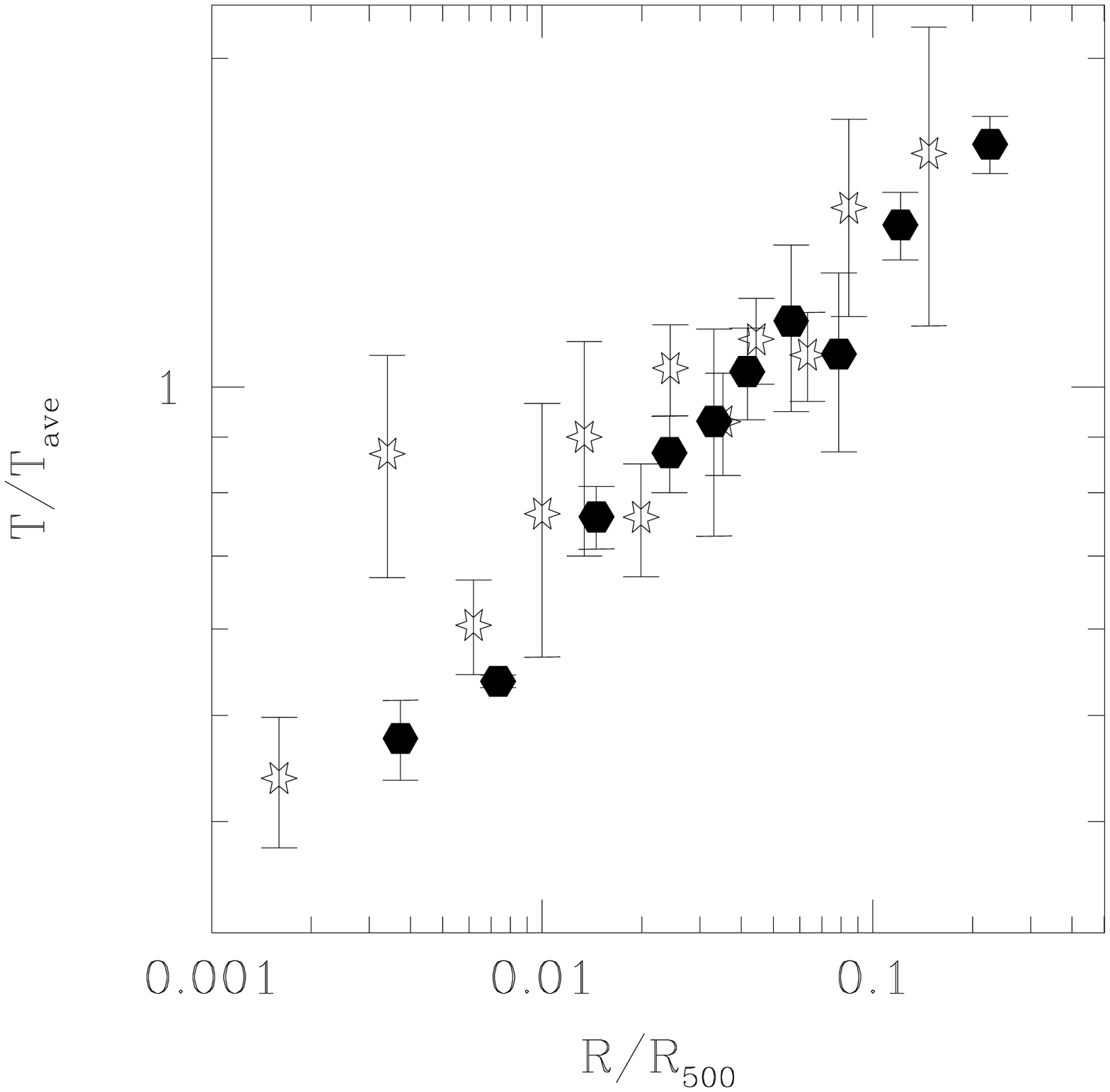}}}
\subfigure{\scalebox{.3}{\includegraphics{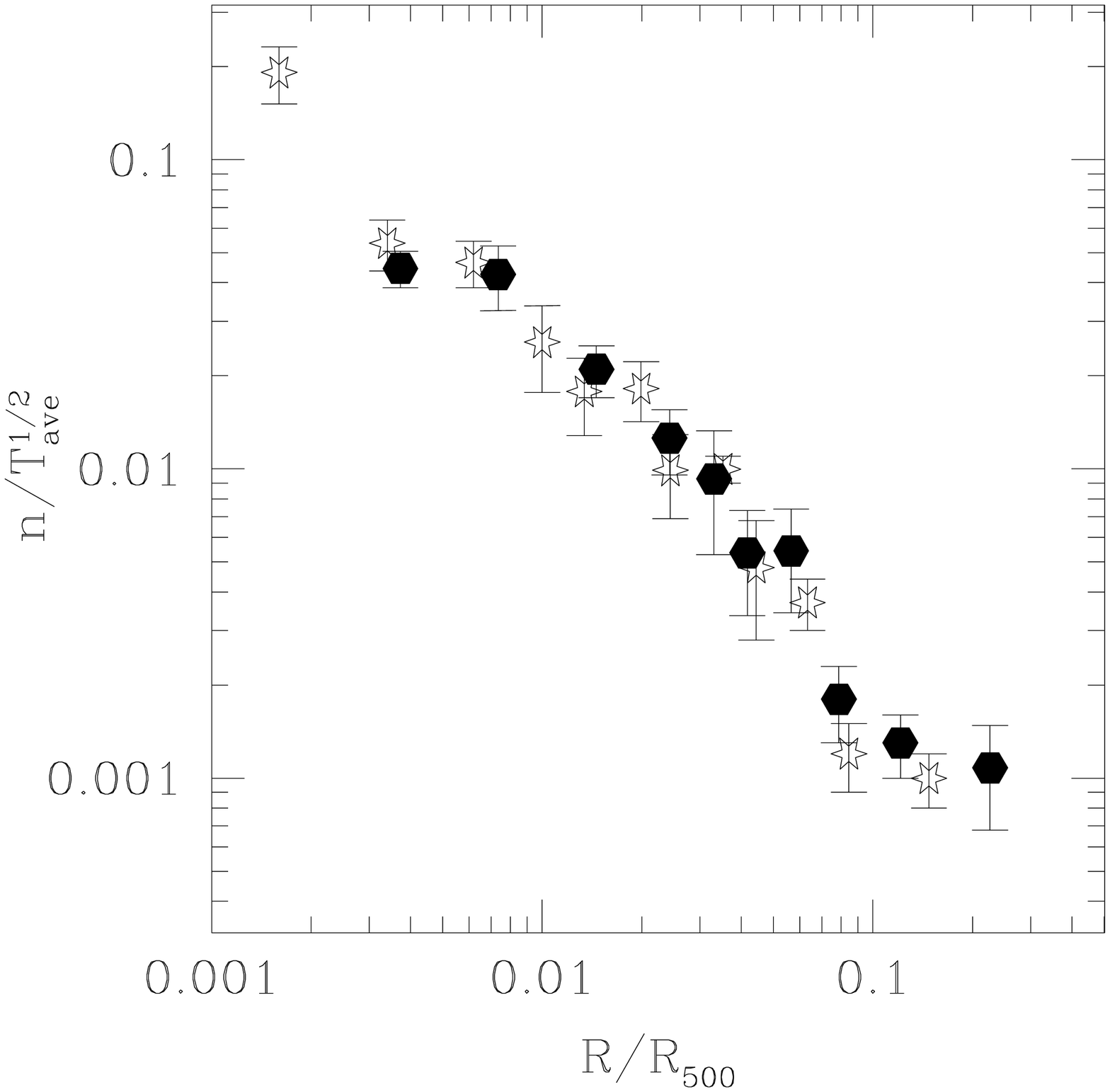}}}
\subfigure{\scalebox{.3}{\includegraphics{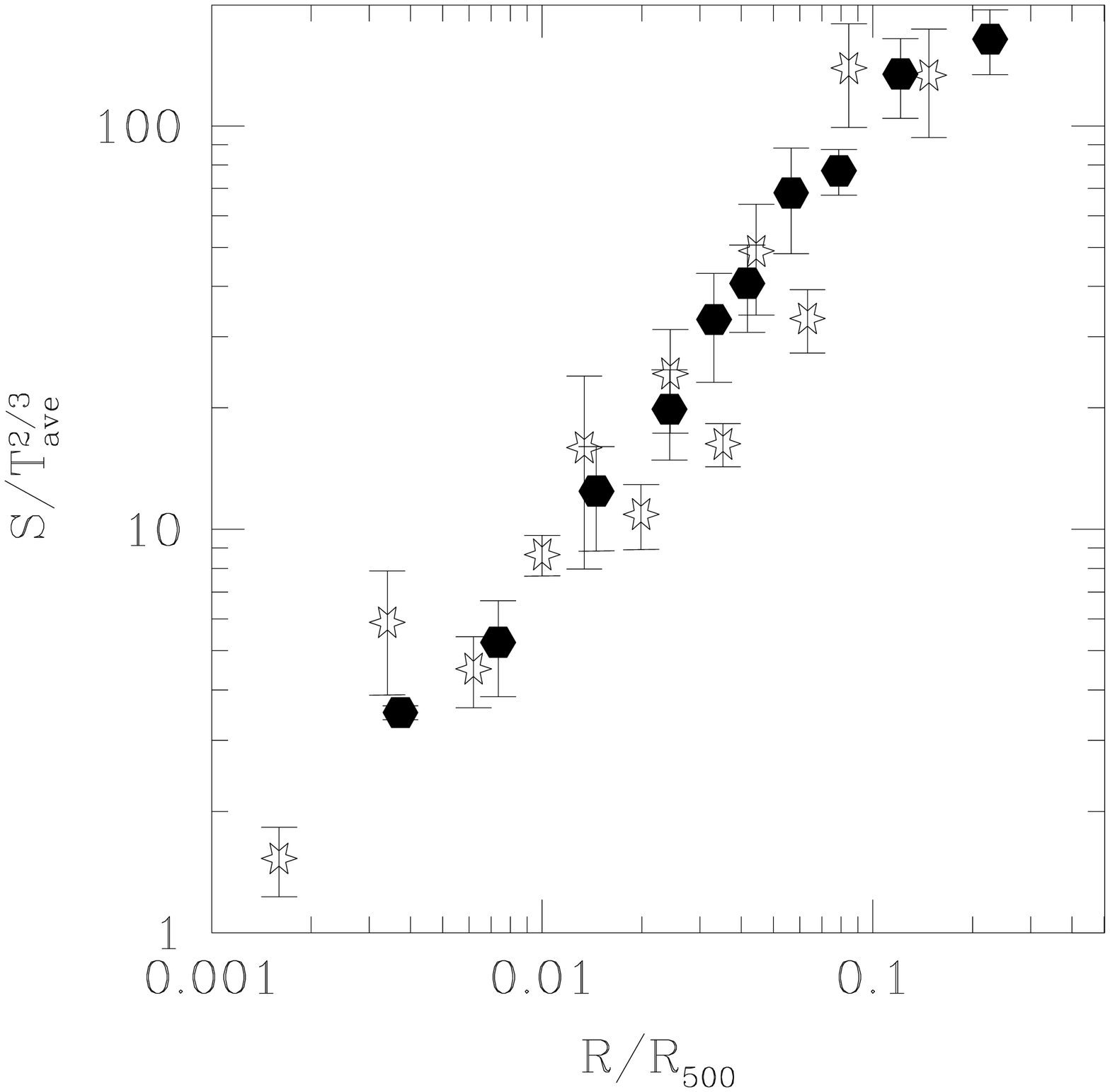}}}
\caption{Co-added, self-similarly scaled temperature, density and entropy profiles (from top
to bottom), of the sample split according to the radio power of the
BGG.  The closed points indicate groups with BGGs such that
$L_{1400}\geq$21.5, and the open points comprise the rest of the
sample.  There is very little difference between the density and
entropy profiles of groups containing radio loud BGGs and those
containing radio quiet BGGs, but that there is a significant
(2$\sigma$) difference between the gradients of the temperature profiles.}
\label{aveprofiles}
\end{figure}

Fitting the subsamples, we find that whilst there are some
differences between the gradients of the scaled temperature and
entropy profiles, the differences are slight, approximately
1-2$\!\sigma$.  Whilst this could indicate a possible difference
between the temperature and entropy profiles of radio loud and radio
quiet groups, we cannot rule out that it is due to statistical scatter.
To test this further, the profiles were co-added and overplotted to
see if the averaged profiles showed any significant differences.  The
averaged profiles are shown in Fig~\ref{aveprofiles} for the
$L_{1400}=$21.5 cut, and the results of the fits for all three cuts
are shown in Table~\ref{results_fitsAGN}. It can be seen that our
data show that there is approximately a $2\sigma$ difference between
the temperature gradients of the radio loud sample compared to the
radio quiet sample (a gradient of 0.31$\pm$0.01 for the co-added radio
loud sample compared to 0.24$\pm$0.04 for the co-added radio quiet
sample).  {\bf The apparent jump in temperature seen at 0.004$R_{500}$
in the radio quiet sample does not seem to be a feature associated
with all radio-quiet groups, but rather primarily the result of one
anomalous system, namely NGC~3607.}  No significant difference in gradients is seen between the
co-added scaled radio loud and radio quiet entropy profiles (gradients
of 1.04$\pm$0.05 for the radio loud sample compared to 1.00$\pm$0.08
for the radio quiet sample).  Further, the radio quiet and radio loud
density profiles show little difference in gradient ($-1.16\pm 0.07$
for the radio quiet compared to $-1.09\pm 0.09$ for the radio loud samples).

\subsection{Black hole mass and heating}
\label{blackholemass}
As only a small difference is seen between radio quiet and radio loud systems,
it could be that the effect of an AGN is cumulative, building up over
repeated duty cycles. If this is so, then comparing profiles of groups
with more massive central black holes with those groups with smaller
black holes could provide some insight, since a larger black hole
should theoretically have had more accretion cycles than a smaller
one, and thus, more cycles of radio activity that should have injected
more energy into the group.

The relationship between black hole mass and galactic bulge mass has
been investigated by many authors (for example
\citealt{2003ApJ...589L..21M}), and whilst such a
relationship does exist, it contains significant amounts of scatter,
thus we must exercise caution in interpreting correlations that make
use of it.

\begin{figure}
\subfigure{\scalebox{.3}{\includegraphics{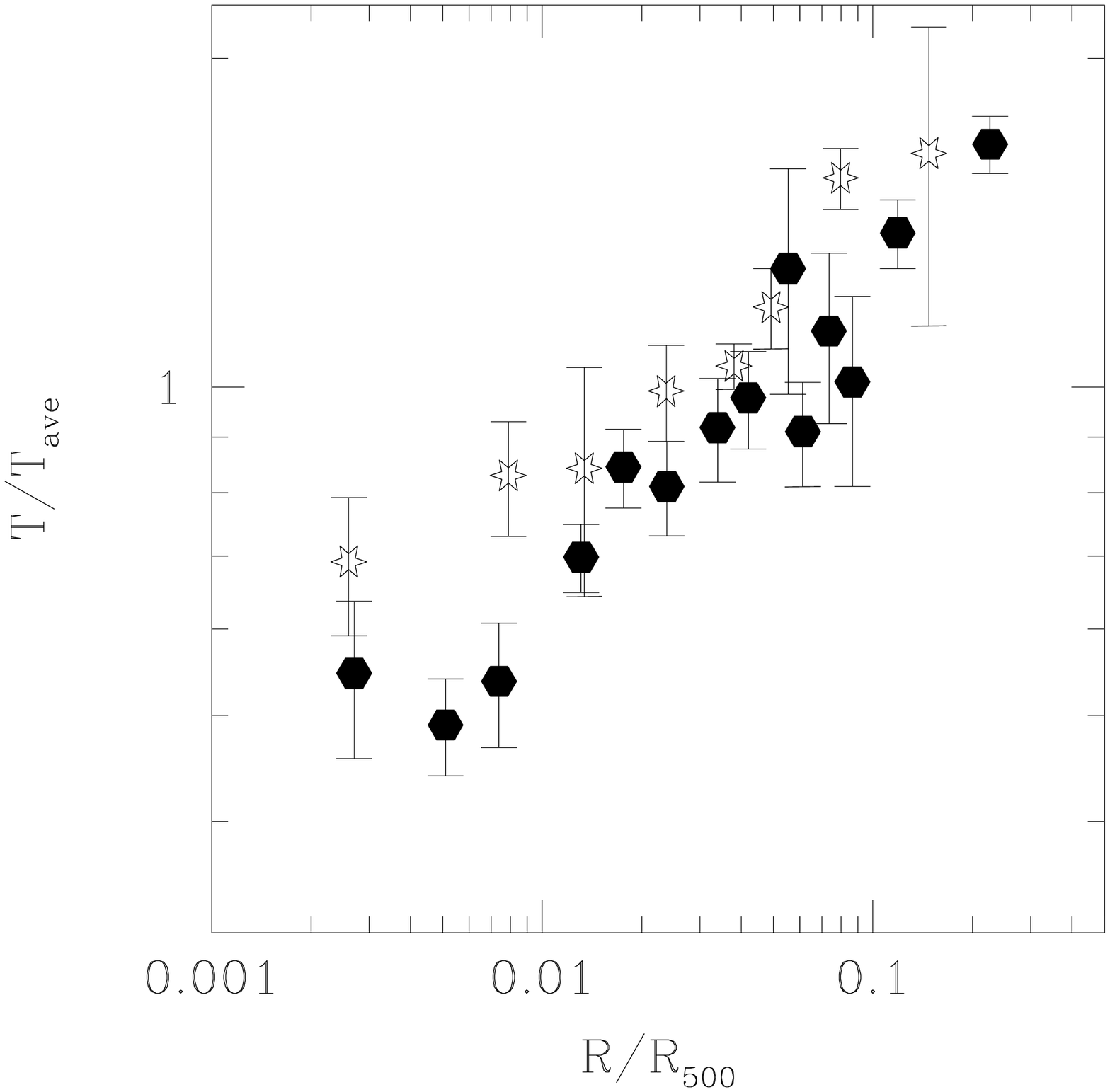}}}
\subfigure{\scalebox{.3}{\includegraphics{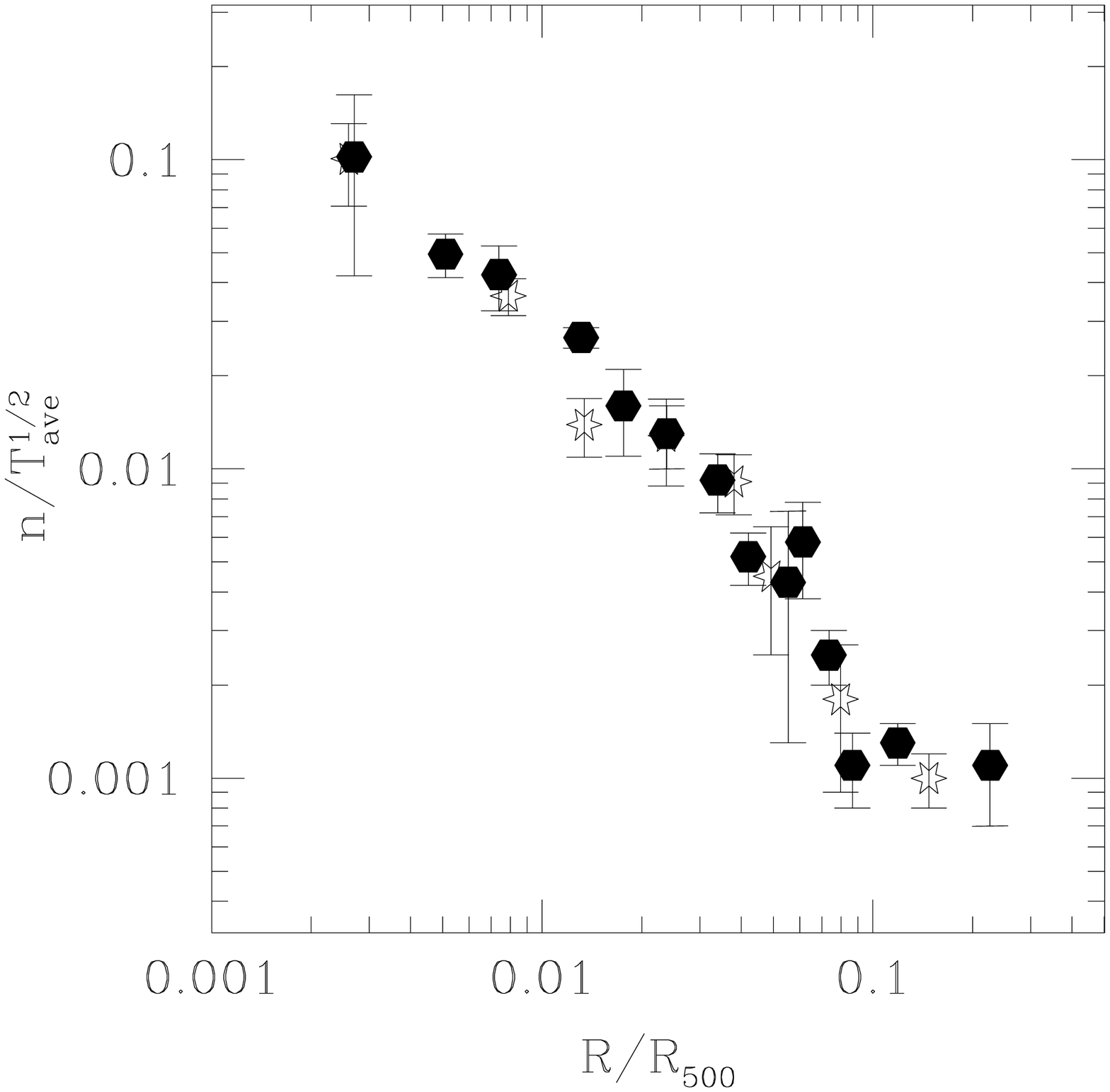}}}
\subfigure{\scalebox{.3}{\includegraphics{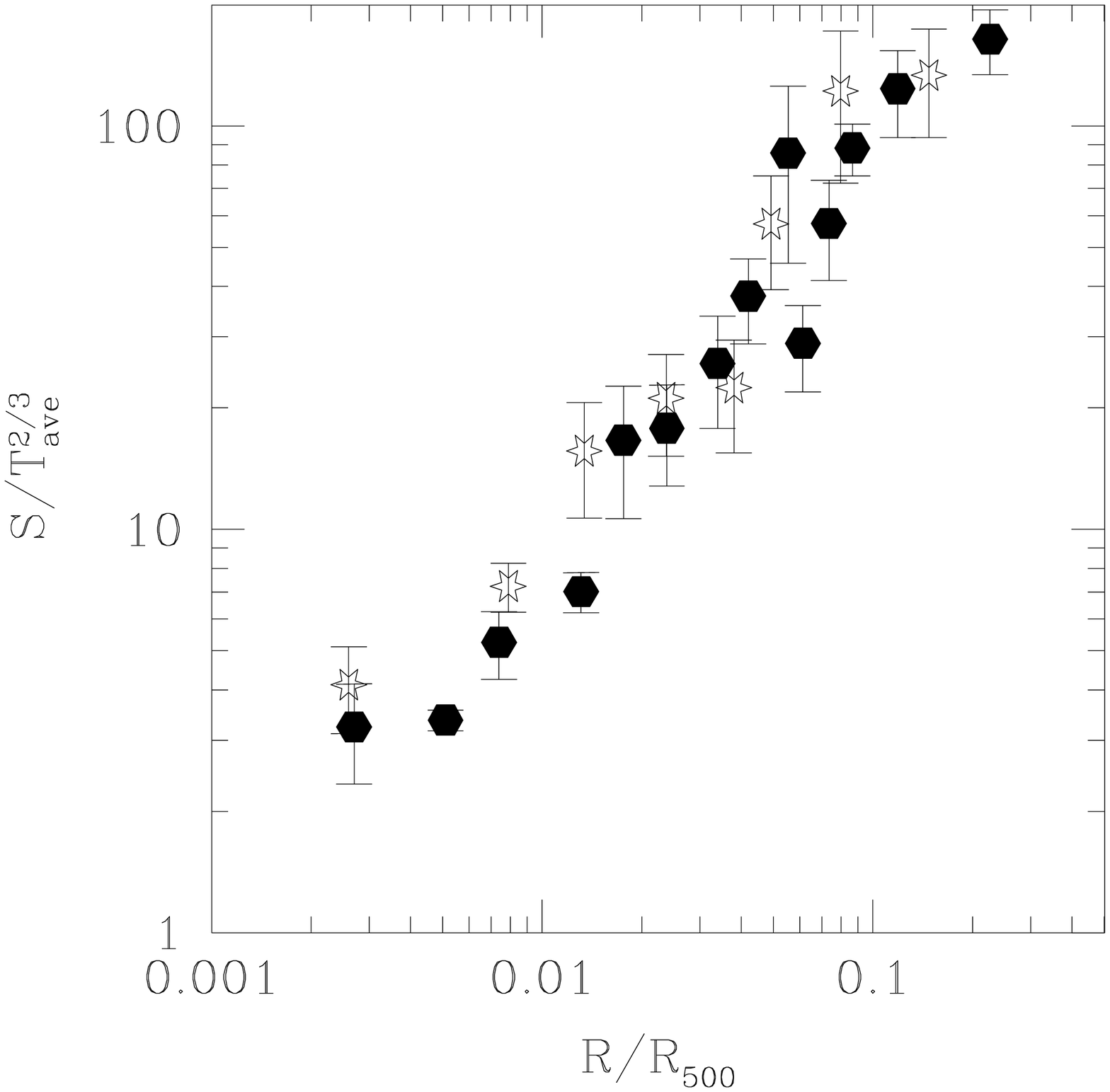}}}
\caption{Co-added, scaled temperature, density and entropy profiles
(from top to bottom) of the sample split into two subsamples, based on
$M_K$.  The solid points represent the subsample for which $M_K <$-25,
whereas the open points represent the subsample where $M_K\geq$-25.
There appears to be a difference between the temperature
profiles of the two subsamples.}
\label{binned_profsBH}
\end{figure}

We obtain K-band luminosities ($M_K$) for the BGGs in our sample from
2MASS (see also Table~\ref{physprops}), and using $M_K$ as a proxy for black hole mass, we divide our
sample into two based on the $M_K$ of the BGG, and create subsamples
for $M_K<-25$ and $M_K\geq -25$.  We co-add the scaled temperature,
density, and entropy profiles for the subsamples, and fit them as in
Section~\ref{AGNsplit}.  The results of fitting the scaled profiles of
the two subsamples are shown in Table~\ref{fit_resultsBH}, and the
co-added profiles are presented in Fig~\ref{binned_profsBH}.

From the plots it can be seen that there is a difference in the scaled
temperature profiles in that groups which host larger BGGs (hence
those BGGs which should have more massive black holes and have
undergone more cycles of accretion) have a lower scaled temperature
within 0.04$R_{500}$ than groups with smaller BGGs.  This could arise
from both subsamples having the same central temperature but with the
$L_K$ bright subsample having an higher overall group temperature.
Fitting the binned profiles, we find that the temperature gradient for
the sample containing groups with larger BGGs is steeper than the
sample of those groups with smaller BGGS (0.28$\pm$0.03 compared with
0.22$\pm$0.02).  From Table~\ref{fit_resultsBH} it can be seem that
there is a small (approximately 1.5$\sigma$) difference between the
entropy profiles of the two subsamples; the fits to the binned entropy
profiles show approximately a 1$\sigma$ difference between the slopes
of the entropy profiles (1.04$\pm$0.07 for the sample with $M_K<-25$
and 0.96$\pm$0.08 for the sample with $M_K\geq-25$).  This could be a
result of a simple correlation between the X-ray scaled gradients of
the hot gas and the BGG mass, since larger BGGs would have steeper
underlying potential wells which would lead to steeper temperature
gradients if the X-ray gas is a tracer of the dark matter.  Further,
we cannot rule out that any correlation between our scaled
temperature, density and entropy profiles may be a result of the size
of the host galaxy rather than black hole mass.

\begin{table*}
\caption{Fits to our sample split according to the mass of the central
black hole.}
\begin{tabular}{lllllll}
\hline Sub-sample        & \multicolumn{2}{c}{Temperature}  & \multicolumn{2}{c}{Density}     & \multicolumn{2}{c}{Entropy}\\
              & gradient & intercept             & gradient &
intercept     &  gradient & intercept\\ \hline
$M_K<-25$     &0.28$\pm$0.03&0.43$\pm$0.07&-1.30$\pm$0.07&-4.1$\pm$0.1&1.04$\pm$0.07&3.1$\pm$0.1\\
$M_K\geq-25$
&0.22$\pm$0.02&0.36$\pm$0.06&-1.30$\pm$0.08&-4.1$\pm$0.1&0.96$\pm$0.08&3.1$\pm$0.1
\\ \hline
\end{tabular}
\vspace{0.2cm}
\begin{minipage}{16cm}
\small NOTES: The sub-samples were fitted in logarithmic space using
an orthogonal regression method. Errors are 1$\sigma$ for two
interesting parameters.
\end{minipage}
\label{fit_resultsBH}
\end{table*}

\subsection{Using an alternative scaling}
\label{altscaling}

In Sections~\ref{AGNsplit} and \ref{blackholemass}, the gas profiles
have been scaled by an average temperature obtained from the \ROSAT
data.  As we expect the \Chandra temperature profiles to agree with
the \ROSAT profiles at large distances from the centre of the group,
scaling the Chandra profiles using the \ROSAT temperature almost
guarantees convergence of scaled profiles outside the cores of groups.
An independent temperature scaling would be desirable, to allow an
absolute normalisation of profiles and avoid introducing an artificial
convergence.  To this end, we investigate the use of the velocity
dispersion of the galaxy group, $\sigma_{G}$ as a proxy of the virial
temperature $T_V$.  Since $T_V\propto \sigma_{G}^2$, scaling the
entropy profiles by $\sigma_{G}^{4/3}$ should be equivalent to scaling
by $T_V^{2/3}$.

\begin{figure}
\subfigure{\scalebox{.3}{\includegraphics{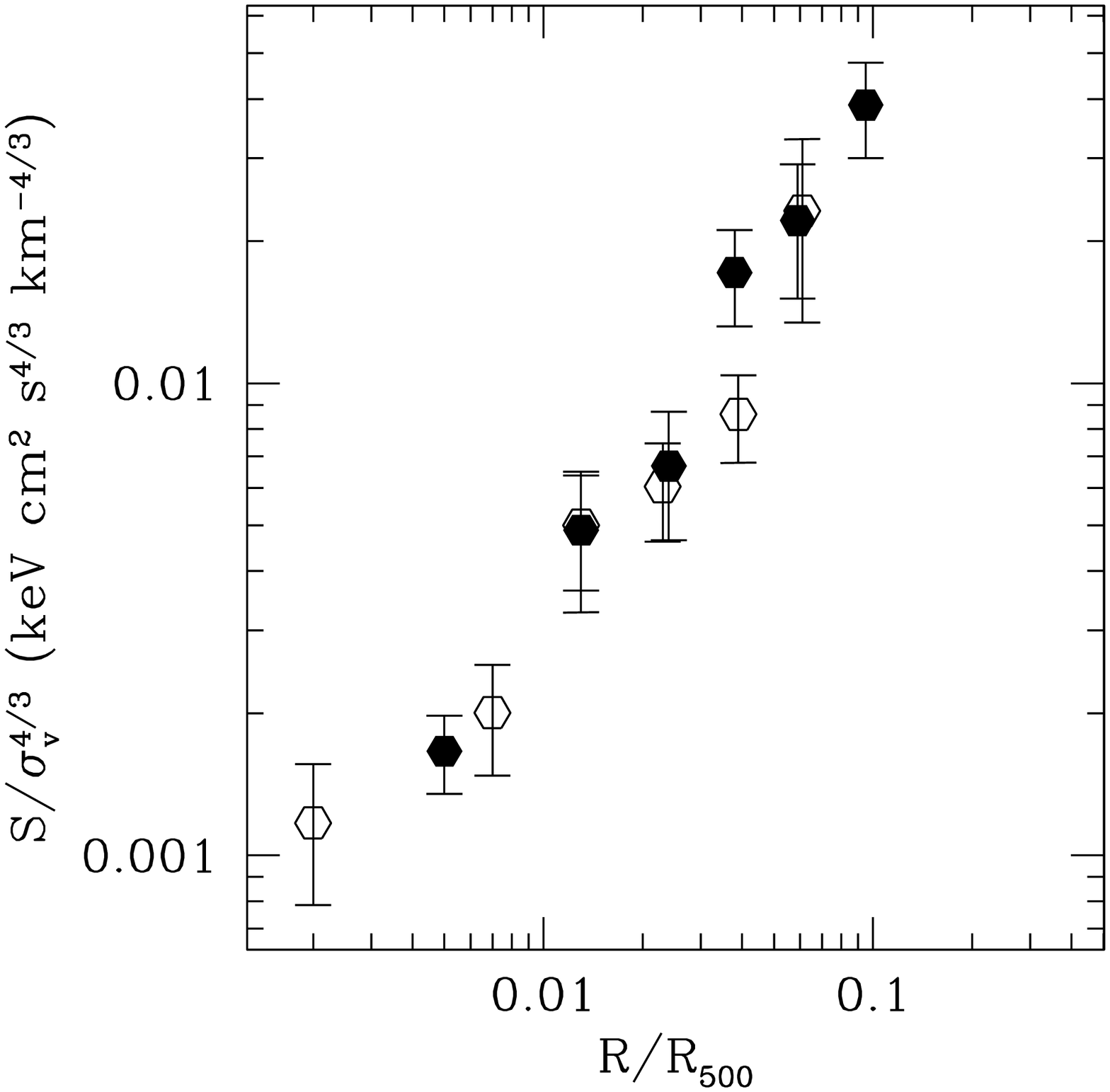}}}\\
\subfigure{\scalebox{.3}{\includegraphics{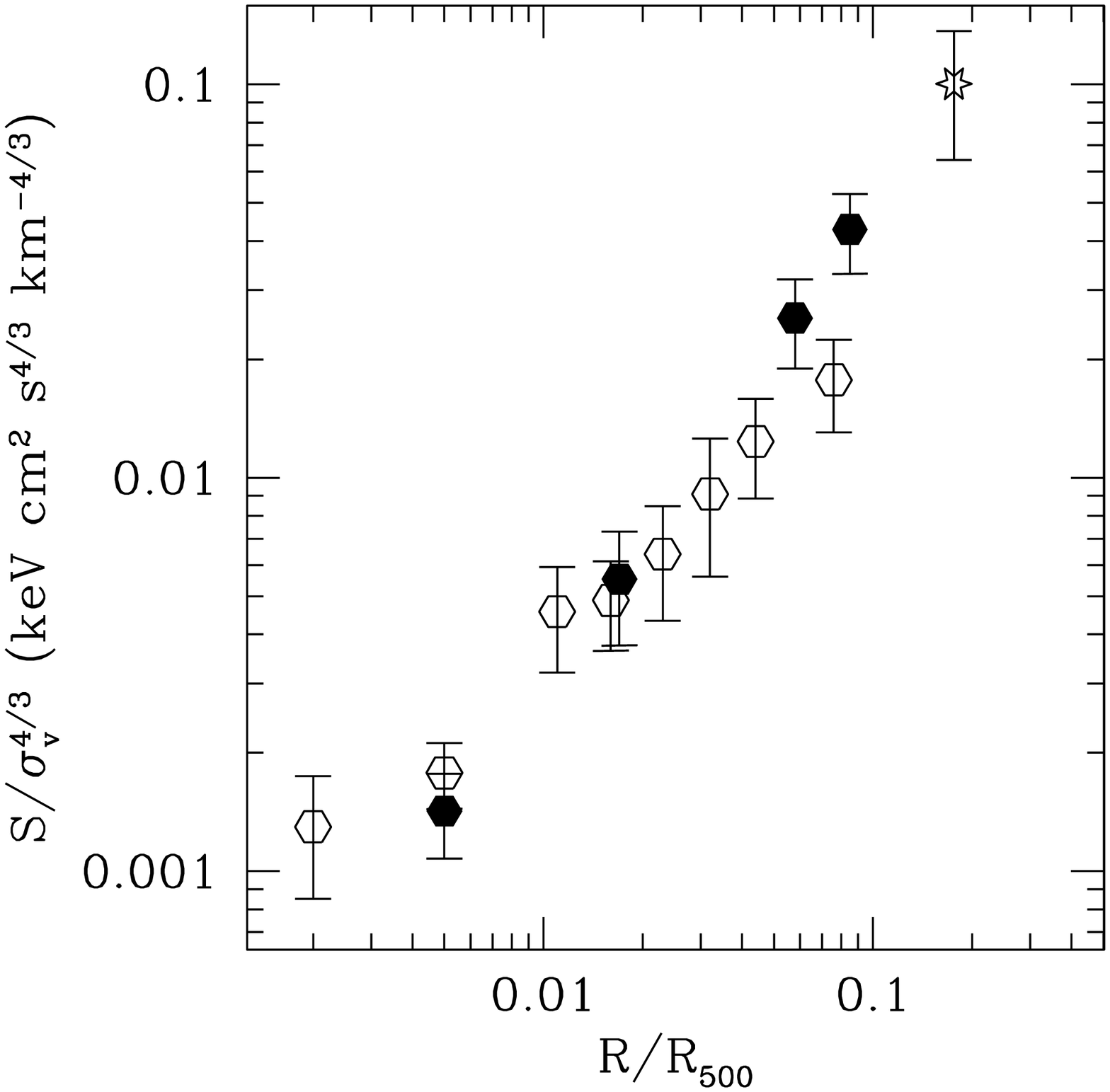}}}
\caption{Entropy profiles for the sample split and binned according to
$\log\left(L_{1400}\right)$ (top) and the galaxy velocity dispersion,
$\sigma$, (bottom).  {\bf In the top plot, the open points represent those
groups in which $\log\left(L_{1400}\right)<21.5$ and the black points
are those where the BGG radio power is greater than this limit.  In
the bottom plot, the open points represent those systems where
$\sigma<270~\mathrm{km\ s^{-1}}$, and the black points are the systems
where $\sigma\geq 270~\mathrm{km\ s^{-1}}$.}  The profiles are scaled
radially by $R_{500}$ and vertically by $\sigma_{G}^{4/3}$.  We see
very little difference between the entropy profiles of radio loud
versus radio quiet groups, whilst groups whose BGGs have large black
holes have steeper entropy profiles than those groups whose BGGs have
smaller black holes.}
\label{sigmasplit}
\end{figure}

We can also improve on the use of $M_K$ as a proxy for the mass of the
central black hole. The $M_K$-black hole mass ($M_{BH}$) relation
shows substantial scatter, whilst the $\sigma:M_{BH}$ relation, where
$\sigma$ is the velocity dispersion of the BGG, is tighter
(\citealt{2000ApJ...539L...9F} and
\citealt{2000ApJ...539L..13G}). {\bf  However, recent work
\citep{2006astro.ph..6739L} suggests that for BGGs, the
$\sigma:M_{BH}$ relation may not be as well constrained compared to
the $M_V:M_{BH}$ relation, where $M_V$ is the V-band magnitude of the
BGG.  This is however a much debated issue, and it has been suggested
that the discrepancy between $M_{BH}$ as predicted by the two
different methods arises due to $M_V$ being overestimated in BGGs
\citep{2006astro.ph.10264B}.  They argue that in fact when black hole
masses obtained using a $\sigma:M_{BH}$ relation are compared to those
using a $M_K:\sigma$ relation, i.e. a near-infra-red rather than blue
magnitude relation, \citep[e.g.][]{2003ApJ...589L..21M}, there is no
discrepancy, and that the discrepancy arises due to the unusual colour
profiles of BGGs compared to E and S0 galaxies.}  With this in mind,
and with no apparant resolution, we opt to use the tighter
$\sigma:M_{BH}$ relation, and scale the entropy profiles by $R_{500}$
radially, and by $\sigma_{G}^{4/3}$ vertically, and split the sample
into two in two different ways; in Fig~\ref{sigmasplit}a, as in
Section~\ref{AGNsplit} at $\log \left(L_{1400}\right)=21.5$ and in
Fig~\ref{sigmasplit}b at $\sigma=\mathrm{270\ km \ s^{-1}}$.  Values
for $\sigma$ are collated from
Hyperleda\footnote{http://leda.univ-lyonl.fr}, and shown in
Table~\ref{physprops}.  The subsamples for both the
$\log\left(L_{1400}\right)$ and $\sigma$ splits are fitted as before.
We find that the entropy profiles in the $\log\left(L_{1400}\right)$
split show no significant differences between the profiles, whilst the
$\sigma$ split shows that systems with higher values of $\sigma$ have
steeper entropy profiles than those with lower values (gradients of
1.22$\pm$0.01 compared to 0.78$\pm$0.05), suggesting that the effect
seen in Section~\ref{blackholemass} is real.

\subsection{Effects of AGN activity on individual systems}
\label{indsys}

\begin{figure}
\subfigure{\scalebox{.3}{\includegraphics{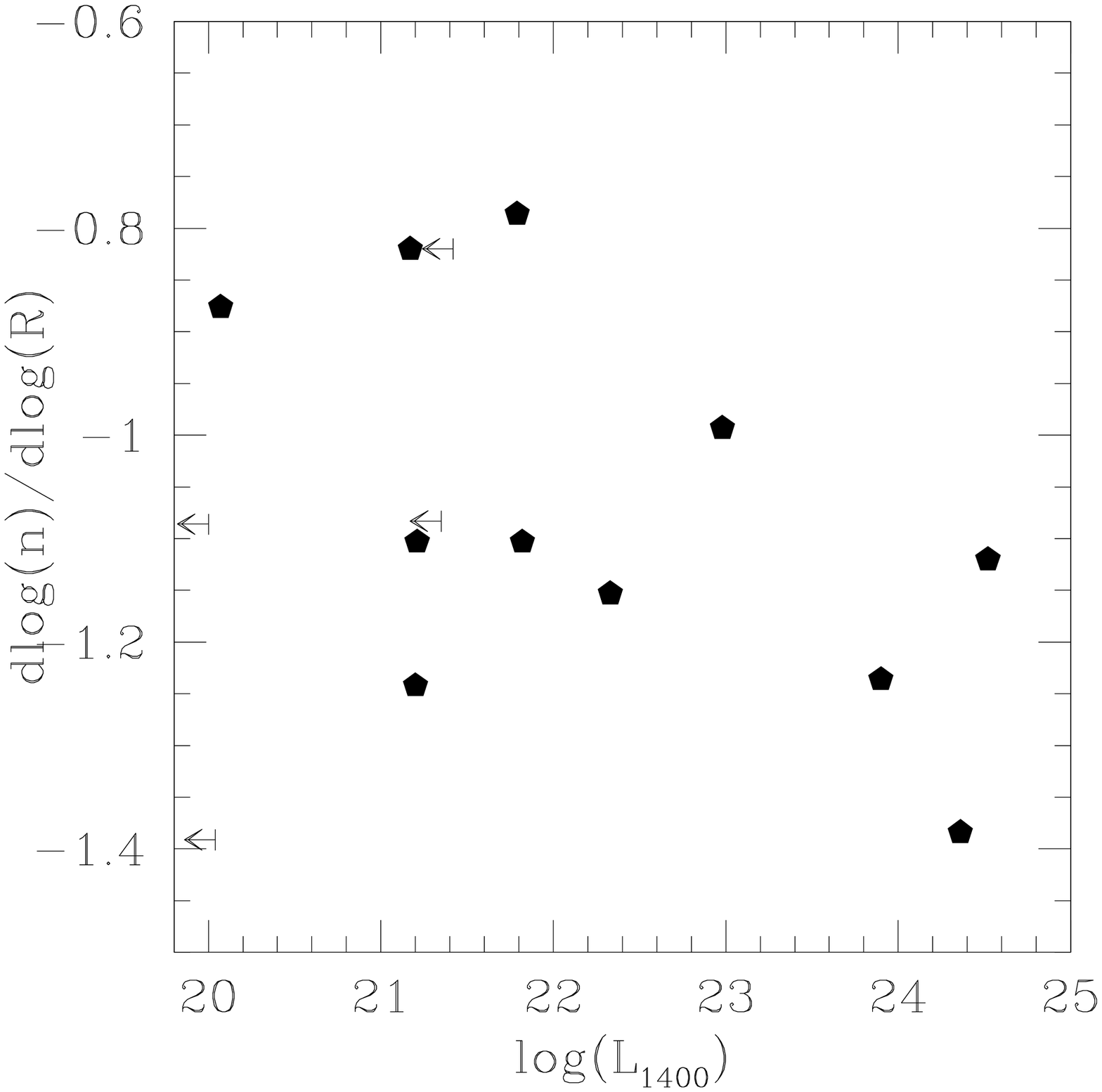}}}
\subfigure{\scalebox{.3}{\includegraphics{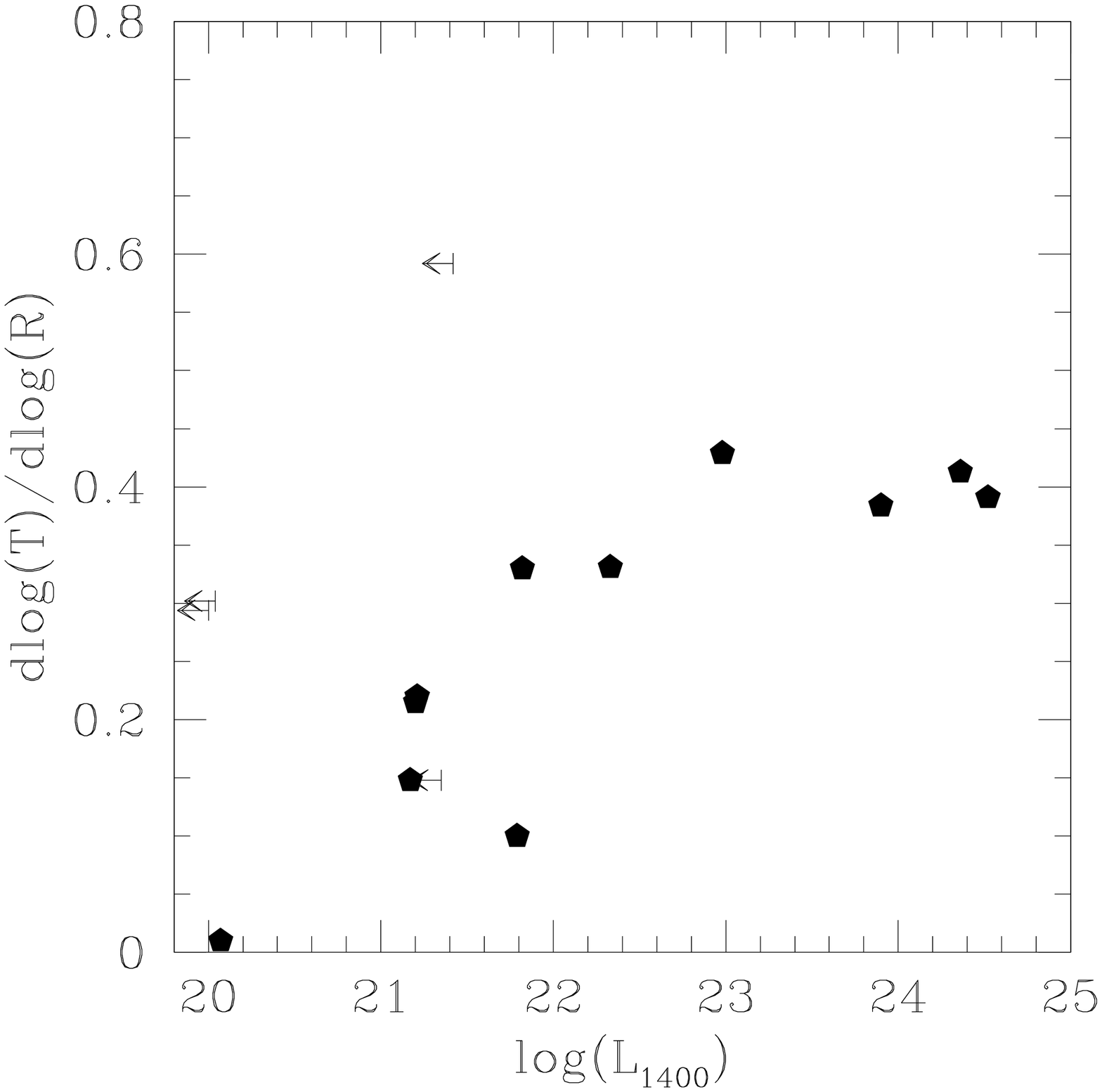}}}
\subfigure{\scalebox{.3}{\includegraphics{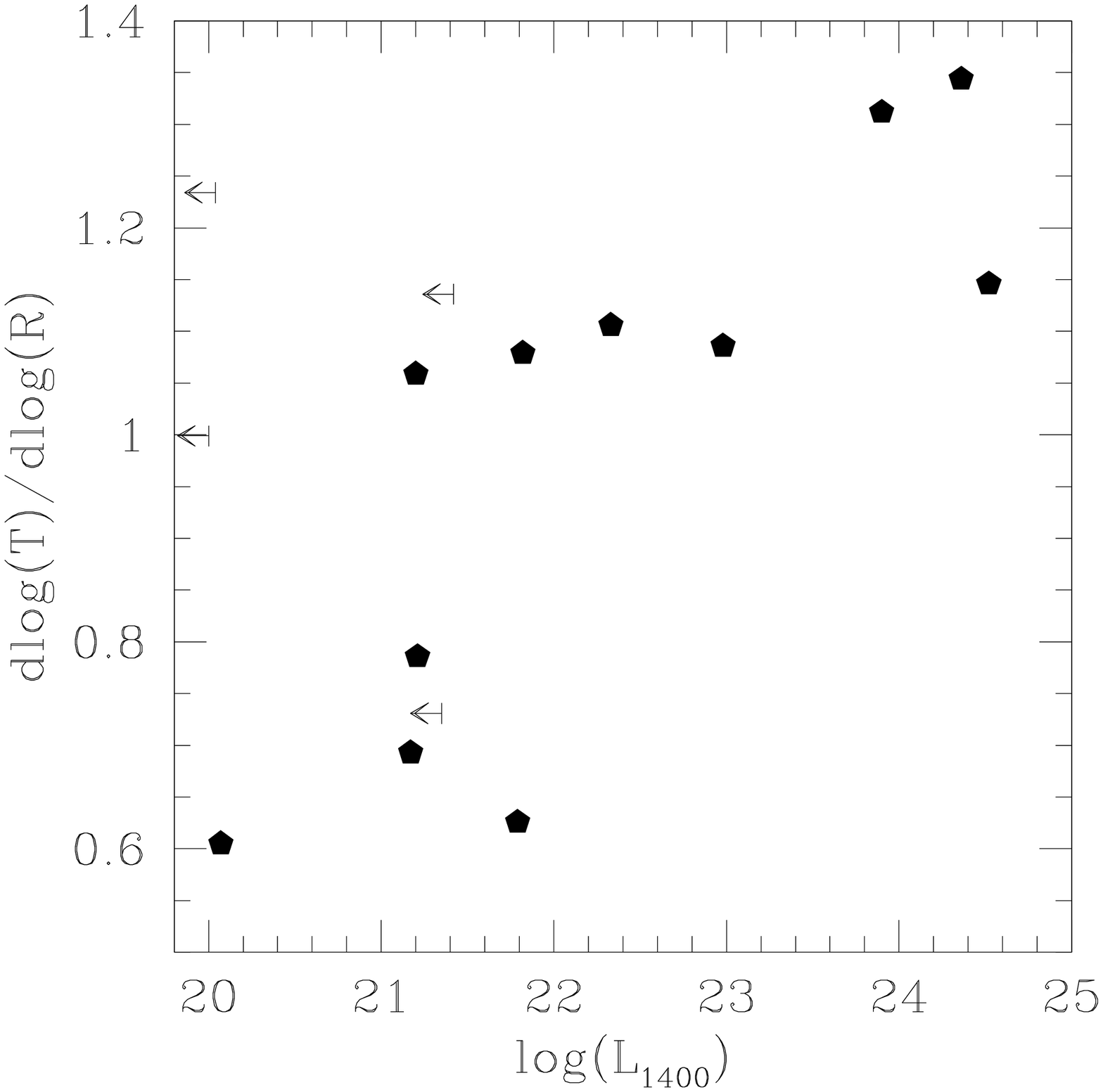}}}
\caption{The logarithmic density, temperature and entropy gradients
plotted against $L_{1400}$.  There exists some correlation between the
radio power and the entropy and temperature gradients, but not with
the density gradient.}
\label{plotsperr}
\end{figure}

As has been discussed in Sections~\ref{AGNsplit} and
\ref{blackholemass}, when comparing the sample as a whole, there
appears to be some marginal difference between the gas properties of
the groups with radio loud BGGs and those with radio quiet BGGs.  The
fact that the difference is marginal could indicate that a real effect
is being diluted by averaging together groups with different radio
properties.  To test this we correlate the logarithmic gradients of
scaled temperature, entropy and density profiles of individual groups
with $\log\left(L_{1400}\right)$ of the BGG. The Pearson $R$
correlation coefficients are shown in Table~\ref{pearsonr}, and plots
of the various quantities that have been correlated are shown in
Fig.~\ref{plotsperr}.  We find significant
correlations between $L_{1400}$ and both the temperature and entropy
gradients.

Since our sample contains three systems with extensive jet-fed radio
lobes, it is interesting to examine whether these systems might be
entirely responsible for the rather weak correlations we see.  These
three systems (NGC~383, NGC~741, NGC~4261)  have jets which transport the AGN
energy out to large radii where the gas density may be too low to
allow the energy to be radiated in less than a Hubble time. This could
lead to steepening of temperature and entropy profiles in these
systems.  We find that if the three groups with large radio galaxies
are excluded from the sample the correlations are no longer
significant. Hence our data do not require any effect on the gas from
AGN beyond that arising in the three large jet systems.

\begin{figure}
\subfigure{\scalebox{.3}{\includegraphics{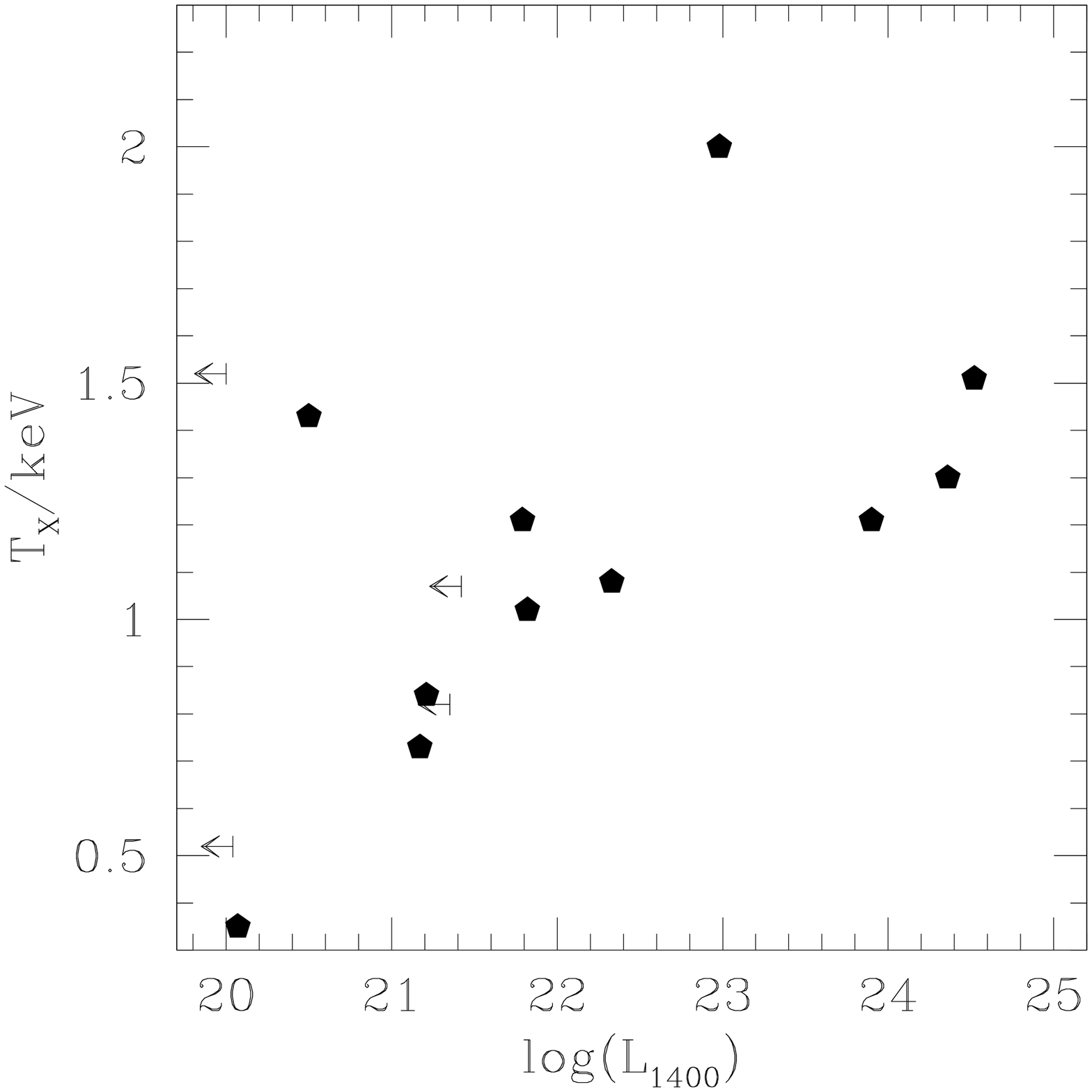}}}
\subfigure{\scalebox{.3}{\includegraphics{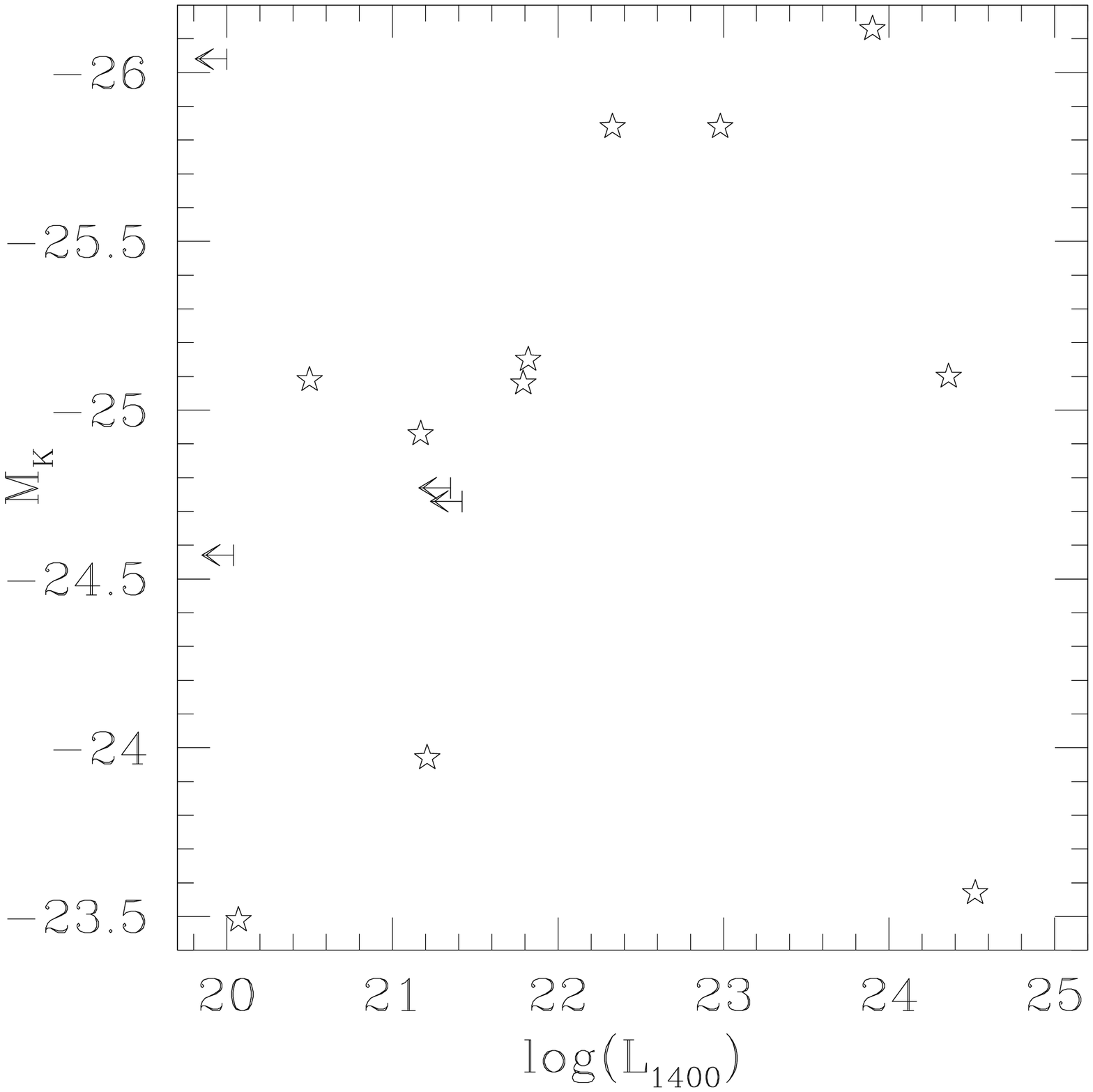}}}
\subfigure{\scalebox{.3}{\includegraphics{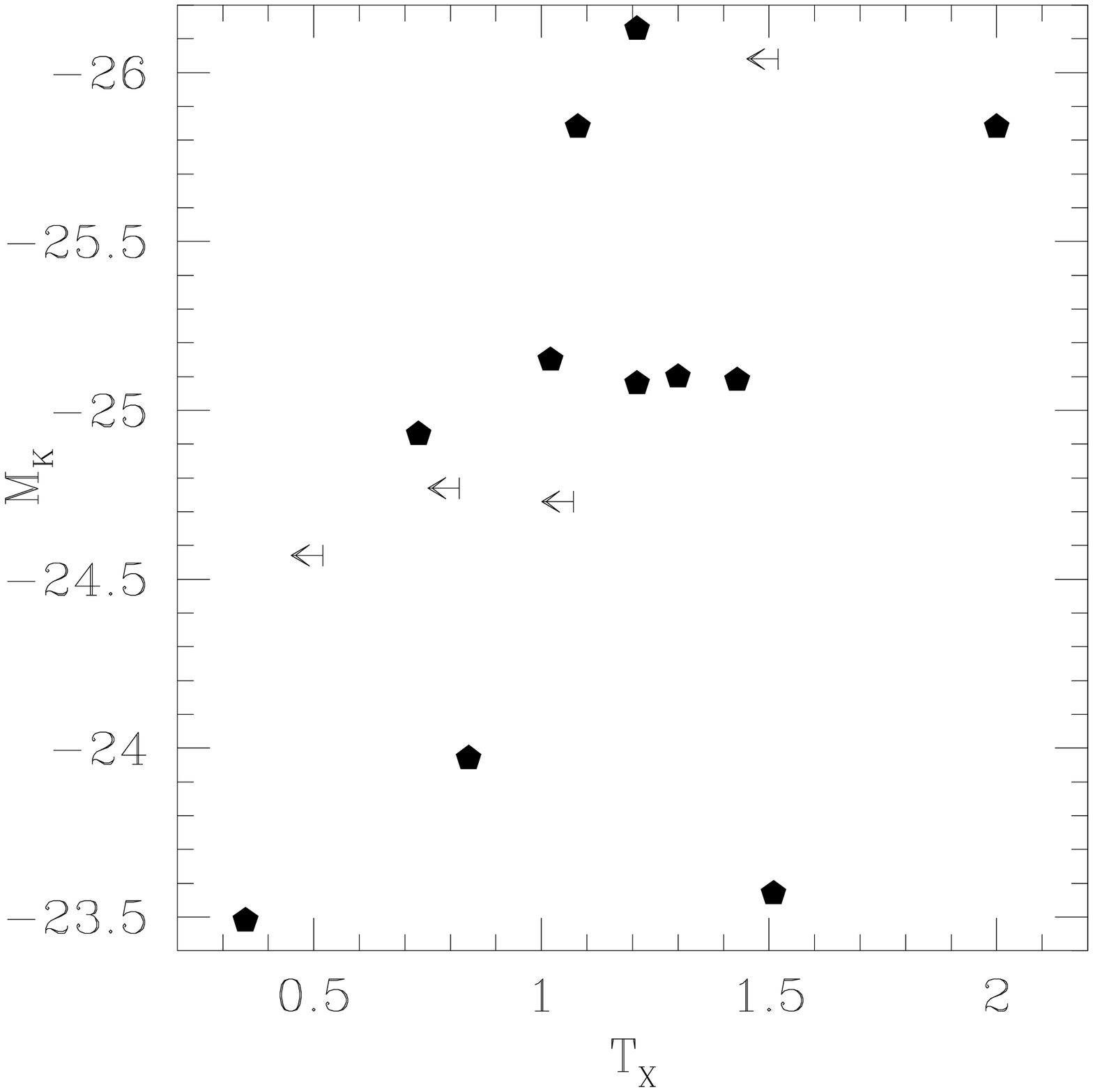}}}
\caption{Plots of $T_X$ against $L_{1400}$ (top), $M_K$ against
$L_{1400}$ (middle) and $T_X$ against $M_K$ (bottom).  Only $T_X$ and
$L_{1400}$ are significantly correlated.}
\label{corelfig}
\end{figure}

We further correlate $L_{1400}$ of the BGG, with the global X-ray
temperature, $T_X$, as measured by \ROSAT, and with $M_K$, the K-band
magnitude of the BGG (Table~\ref{corels} and Fig.~\ref{corelfig}).  We
also find that $L_{1400}$ correlates with $T_X$.  This correlation
suggests that either the radio source does heat the IGM, or
alternatively, that a hotter group, with a strong temperature and
entropy gradient is more conducive to triggering an AGN outflow of the
scale of the largest radio galaxies, than a cooler group.  There
appears to be no significant correlation between $L_{1400}$ and $M_K$
or $M_K$ and $T_X$, indicating that the effect is not due to larger
galaxies hosting larger radio outbursts and being found in hotter
groups.  This suggests that the difference in temperature profiles found
in Section~\ref{blackholemass} may arise from steeper potential wells
in the systems with larger BGGs.

\begin{table}
\caption{Correlation coefficients for logarithmic temperature, entropy
and density gradients and $\log\left(L_{1400}\right)$.  The final column indicates
whether the correlation is significant at the 95\% confidence level,
i.e. can the null hypothesis (that there is no correlation) be
rejected with only a 5\% level of uncertainty.} 
\begin{tabular}{lll}
\hline Quantity & R    & Significant?\\ \hline
$d \log T / d \log R$  & 0.50 & Y$^1$\\
$d \log S / d \log R$  & 0.56 & Y$^1$\\
$d \log n / d \log R$  &-0.23 & N\\ \hline
\end{tabular}
\vspace{0.2cm}
\begin{minipage}{8cm}
\small NOTES: $^1$The correlation disappears when the sources with
powerful radio jets are removed from the sample (see text).
\end{minipage}
\label{pearsonr}
\end{table}

\begin{table}
\label{corels}
\caption{Correlation coefficients for radio luminosity, X-ray
temperature, and $M_B$ of the systems in our sample.  Correlations marked
with $^1$ are those for which a correlation significant at 95\% was found.}
\begin{tabular}{l|lll}
Quantity  & $L_{1400}$ & $T_X$     & $M_K$ \\ \hline
$L_{1400}$&    -       & 0.51$^1$   &-0.14  \\
$T_X$     &    -       &  -        &0.46  \\ \hline
\end{tabular}
\end{table}

\subsection{The $L_X\!:T_X$ relation}
\label{LxTx}
\citet{2005MNRAS.357..279C} find that groups which host radio galaxies
appear to fall below the $L_X:T_X$ relation for groups
(c.f. \citet{OP2004}), which they interpret as evidence for AGN
heating of galaxy groups.  We use the temperatures and luminosities
extracted in Section~\ref{obtaininglt} to produce an $L_X:T_X$
relation for the inner regions of groups as sampled by the \Chandra
data; the corresponding $L_X:T_X$ relation is shown in
Fig.~\ref{LTrel}a.  

From Fig.~\ref{LTrel}, there does appear to be some difference in the
distribution of radio loud and radio quiet points around the mean
$L_X:T_X$ relation, and to test this, we measure the perpendicular
distance for each group from the mean $L_X:T_X$ fit, and as above,
compare the distribution of distances for the radio loud and radio
quiet systems.  A K-S test indicates that
there is a 20\% chance that the observed scatter around the mean
relation would arise by chance if the radio loud and radio quiet
sub-samples came from the same population ($D$=0.46, $P$=0.20),
implying that the effect is marginally significant.  However, we are
dealing with only 19 systems, and so both effects need to be explored
with a larger sample.

If the difference in distributions around the mean relation is real,
then fixing the gradient of the relation to that obtained from the fit
to the entire sample allows an estimate of the temperature offset
between the radio loud sample and the radio quiet sample to be
determined.  We find that the temperature offset $\Delta T =0.2\pm
0.1$~keV.  If this temperature offset is entirely due to the impact
of AGN heating within $0.05R_{500}$, then an estimate of the energy
required to produce this offset can be obtained.

The energy input $\Delta E$ required to produce a temperature change
of $\Delta T$ is given by:
\begin{equation}\Delta E = Nk\Delta T\end{equation} where $N$ is the
total number of particles within $0.05R_{500}$ and $k$ is the
Boltzmann constant.  Assuming that the average radio quiet profile is roughly
equivalent to an unheated group, we use the density profile shown in
Fig~\ref{aveprofiles}b, and a $\Delta T=0.2$~keV to estimate $\Delta
E$ for a 1~keV system.  We find that $\Delta E \sim 2.6\times
10^{57}$~erg.  Assuming an AGN lifetime of $10^7$~yrs, this gives an
energy injection rate of $\sim 8\times 10^{42}~\mathrm{erg\ s^{-1}}$.  We
comment below on the plausibility of this injection rate.

We further compare the marginal effect found in the \Chandra data with
the result of \citet{2005MNRAS.357..279C}, to ascertain whether any
AGN heating is taking place within the core, compared to outside the
core.  To do this, we examine the mean residuals ($\sigma T$) of the
radio loud and radio quiet subsamples compared to the fit to the radio
quiet sample.  For the \ROSAT sample of \citet{2005MNRAS.357..279C},
we use the best fitting least squares fit to their {\it c}2 radio
quiet sample in order to have a reasonable comparison of radio power
with our $\log\left(L_{1400}\right)=21.5$ cut.  We find that in the
\ROSAT data, the mean residual for the radio quiet group is $\sigma
T_{RQ}$=0.06$\pm$0.1~keV, compared to the value for the radio loud
sample, $\sigma T_{RL}$=1.6$\pm$0.5~keV.  The corresponding quantities
($\Sigma T$) for the \Chandra data are $\Sigma
T_{RQ}$=0.07$\pm$0.1~keV and $\Sigma T_{RL}$=0.2$\pm$0.1~keV.  The
mean residual is much larger for the radio loud sample in the \ROSAT
data than in the \Chandra data, implying that more significant AGN
heating may be taking place outside the region being probed here
($\sim 0.1R_{500}$).

\section{Discussion}
\label{interpretation}

To summarise the key results above, we find modest differences between
the gas properties of the cores of radio loud groups compared to radio
quiet groups; radio loud groups appear to have slightly steeper
temperature profiles, and there are marginally significant
correlations between the temperature and entropy gradient of groups
and the radio power of their BGGs, though this may arise from the
three strong jet sources in our sample.  We also find that when the
sample is split according by a proxy of the black hole mass (either
$M_K$ or $\sigma$), steeper temperature and entropy gradients are
found for those groups whose central galaxies have large values of
$M_K$ or $\sigma$, compared to those groups where $M_K$ or $\sigma$
are lower.  Considering the effect of active radio galaxies on the
scaling relations, we find evidence for a small offset in the
$L_X:T_X$ relation between radio loud and radio quiet groups, but this
is much smaller than the offset seen on larger spatial scales by
\citet{2005MNRAS.357..279C}.  

In light of these results, we now discuss the three possible
explanations for any observed correlations that were highlighted in
Section~\ref{introduction}.

\begin{itemize}
\item {\bf The radio source modifies the X-ray properties through heating
or displacement of the gas.}  From the images presented in
Section~\ref{imaging} and the work of other authors, it is apparent
that the radio source can displace gas, creating cavities in the X-ray
emission.  The entropy profiles when split on radio power imply that
no extra energy is being injected into the core by current radio
source heating, although systems which exhibit powerful jets may be
transporting extra energy to regions outside the core, and causing a
steepening of the entropy gradient outside the core.  However,
splitting the sample using proxies of black hole mass suggests that
there may be a cumulative effect of repeated AGN outbursts.  It is not
clear whether the cumulative effect is from repeated AGN outbursts, or
due to energy injection from the stellar content of the galaxy.

\item {\bf The observed X-ray properties provide suitable conditions for
radio source triggering.}  If this were the case, then it would be
expected that the immediate gaseous environments of currently radio
active sources would be different from the environments of currently
radio quiet sources.  It can be seen from Fig~\ref{aveprofiles} that
the profiles of radio loud and radio quiet groups are rather similar,
and from Figs~\ref{scaledtemp}-\ref{ct} the scatter between individual
profiles is larger than any systematic differences seen in the
averaged profiles.  As any structural differences conducive to the
production of a radio source would have to be most significant in the
core of the group, since that is where the black hole resides, the
similar profiles for the radio loud and radio quiet groups argue
against this hypothesis.

\item {\bf There exists some third parameter that correlates with both radio
source activity and position on the $L_X:T_X$ relation.}  From
Fig~\ref{corelfig}b, it could be the case that $M_K$ may play some
role, as the mean magnitude for the radio loud sample does seem to be
larger than the mean magnitude for the radio quiet sample.  This could
be a result of a cuspier potential caused by a larger central galaxy.
\end{itemize}

The evidence suggests that our first and third hypotheses could,
together, help explain the correlations found here, whilst the second
hypothesis seems unlikely, since any systematic differences between
radio loud and radio quiet groups found are small, and can be
attributed to either statistical scatter, or to a small number of
sources with physically large sizes that transport energy out to
larger radii.  The fact that there are no strong differences between
radio-loud and radio-quiet groups implies that in general radio
sources do not inject enough energy into the central regions of their
host groups to cause observable heating effects.  If there is any
heating going on on these scales, it must just be enough to balance
cooling.

To investigate whether AGN output is indeed sufficient to balance
cooling, we consider the case of NGC~383 (3C\,31).  This source has
been well studied and is one of the few in our sample that has an
estimated total jet power.  We define the cooling radius as the radius
at which the cooling time is equal to the Hubble time, which for
NGC~383 is $\sim 25\ \mathrm{kpc}$.  Within this radius, $L_{bol}\sim
5 \times10^{41}\ \mathrm{erg\ s^{-1}}$.  The total jet energy flux at
12~$\mathrm{kpc}$ is $\sim 9\times 10^{43}\ \mathrm{erg\ s^{-1}}$
\citep{2002MNRAS.336.1161L}.  Further, in Section~\ref{LxTx}, we
calculated that the average temperature offset between the radio loud
sample and the radio quiet sample is $\sim$0.2~keV.  Over $10^{7}$
years, an energy injection rate of $\sim 8\times 10^{42}~\mathrm{erg\
s^{-1}}$ is required to produce the offset in the NGC~383 group.  This
is about an order of magnitude less than the maximum AGN energy input
rate, so that even a less luminous AGN, where the AGN energy flux
would not be as high, could provide sufficient energy to both
counteract radiative cooling, and raise the temperature of the group.
This agrees well with the conclusions of \citet{2002MNRAS.334..182H}.

Further, the large scale radio morphology of NGC~383 and the fact that
the bolometric luminosity of the cool core is much less than the radio
source power implies that an outburst such as that in NGC~383 could transport
much of the energy contained in the radio source out to larger radii,
where it could also affect the hot gas, explaining the
\citet{2005MNRAS.357..279C} results.  However, the radio morphology
of NGC~383 is very different to most of the other radio outbursts in
this sample, and its long jets and plumes may be transporting energy
to regions outside the core of the group without affecting the
innermost regions.  Radio outbursts on smaller scales may affect the
IGM closer to the BGG.
 
\section{Conclusions}
\label{conclusions}

We have presented azimuthally averaged profiles of the properties of
the central hot gas, and an $L_X\!:T_X$ relation {\bf within
$0.05R_{500}$} for 15 galaxy groups with varying degrees of AGN
activity.

The slightly steeper temperature profiles seen in radio loud groups
could be related to the size of the BGG -- a larger BGG would have a
steeper potential well associated with it, and hence a steeper
gradient -- or it could be that systems with radio-loud AGN have
cooled more than those without radio-loud AGN, and it is this cooling
gas that fuels the radio source.  However, our sample size is small,
and the properties of the gas and how they correlate with the AGN
activity need to be investigated on a wider scale using a larger
sample which probes a wider range of temperatures and scales of AGN
activity.  Alternatively, the effect could arise from the contribution
of a few radio sources with large physical sizes.  These sources may
have little effect on the gas close to the core of the group,
depositing their energy at larger radii, at the outside edge of the
regions probed here, causing the gas profiles of their host groups to
appear to be marginally steeper over our radial range.

The steepening of the entropy gradients in systems with larger values
of $M_K$ and $\sigma$ (indicating larger black holes) compared to
those with smaller values of $M_K$ and $\sigma$ could be evidence for
repeated cycles of AGN activity having an impact on the hot gas in
groups.  However, it is unclear whether the effect has to have
occurred via AGN heating, rather than via other forms of energy
injection from the galaxy.  As both AGN and galaxy growth are linked,
disentangling one effect from the other may be difficult, particularly
as AGN life-cycles are also poorly understood.  Work needs to be done
here in both comparing the gas properties of groups at different
stages of their evolution, and on the life-cycles of AGN hosted by
BGGs.

The $L_X\!:T_X$ relations show that any effect that AGN are having on
the gas primarily occurs at larger radii; it could be that within the
region probed by \Chandra$\!$, AGN only act to counteract radiative
cooling. The difference between the radio loud and radio quiet
subsamples is smaller than the difference found by
\citet{2005MNRAS.357..279C}, suggesting that ongoing large scale
outbursts such as those in NGC~383 and NGC~4261 could be transferring
energy out to large radii, and may have a stronger effect outside the
core, whilst the effects of repeated outbursts are more likely to show
up in the core.  However, more X-ray work is needed here to further
probe and constrain the gas properties at larger radii.

We conclude that it appears that radio loud AGN do not irreversibly
raise the entropy in the core of their host galaxy groups.  Rather, it
seems that the feedback mechanisms at work result in typical central
AGN generating just sufficient energy to balance cooling in the core
of the group.  Repeated outbursts may have some longer lasting effect
on the gas, which may contribute to similarity breaking, although
detailed studies extending to larger radii are required to investigate
this. This cumulative effect may take place primarily via rather rare
large outbursts, which dump energy outside the cooling radius via
radio jets.

\section*{Acknowledgements}
The authors would like to thank Irini Sakelliou for her invaluable
help during the early stages of this study.  MJH thanks the Royal
Society for a research fellowship.  We thank the anonymous referee for
their useful comments in the preparation of this paper.  This research
has made use of the NASA/IPAC Extragalactic Database (NED) which is
operated by the Jet Propulsion Laboratory, California Institute of
Technology, under contract with the National Aeronautics and Space
Administration.
\bibliographystyle{mn2e}
\bibliography{MN061095MJ_bibliography}

\label{lastpage}

\end{document}